\pdfoutput=1

\documentclass[aps,prl,twocolumn,amsmath,amssymb,floatfix,longbibliography,superscriptaddress,footinbib]{revtex4-2}

\usepackage[utf8]{inputenc}
\usepackage{newtxtext}
\usepackage{microtype}
\usepackage{textcomp}
\usepackage{eucal}
\usepackage{bm}
\usepackage{siunitx}
\usepackage{comment}

\usepackage{enumerate}
\usepackage{amsfonts}
\usepackage{amsmath}
\usepackage{amssymb}
\usepackage{color}
\usepackage{soul}

\usepackage{graphicx}

\usepackage[colorlinks,allcolors=blue]{hyperref}
\usepackage[capitalize]{cleveref}

\definecolor{Blue}{rgb}{0.3,0,1}
\definecolor{Red}{rgb}{1,0,0}
\definecolor{Black}{rgb}{0,0,0}

\newcommand{\Imag}{{\Im\mathrm{m}}}   
\newcommand{\Real}{{\mathrm{Re}}}   

\newcommand{\ve}{\boldsymbol} 

\newcommand{\be}{\begin{equation}}
\newcommand{\ee}{\end{equation}}

\newcommand\comdot{%
  \mathrel{{\ooalign{\hss\raisebox{-0.3ex}{$,$}\hss\cr\raisebox{0.3ex}{$\cdot$}}}}
}

\newcommand\combullet{%
  \mathrel{{\ooalign{\hss\raisebox{-0.3ex}{$,$}\hss\cr\raisebox{0.3ex}{$\bullet$}}}}
}

\newcommand\comtimes{%
  \mathrel{{\ooalign{\hss\raisebox{-0.3ex}{$,$}\hss\cr\raisebox{0.3ex}{$\times$}}}}
}

\newcommand{\vehsigma}{\hat{\ve{\sigma}}}

\newcommand{\vep}{\ve{p}}
\newcommand{\veq}{\ve{q}}

\newcommand{\hrhot}{\hat{\rho}_{3}}
\newcommand{\vece}{\ve{e}}
\newcommand{\hrhoz}{\hat{\rho}_{0}}
\newcommand{\veR}{\ve{R}}
\newcommand{\ver}{\ve{r}}

\newcommand{\prlsection}[1]{\textit{#1}.\kern0.05em---\kern0.05em\ignorespaces}

\begin{document}
\title{Inverse spin-Hall effect and spin-swapping in spin-split superconductors}
\author{Lina Johnsen Kamra}
\email{linagj@alumni.ntnu.no}
\affiliation{Center for Quantum Spintronics, Department of Physics,\\ Norwegian University of Science and Technology, NO-7491 Trondheim, Norway}
\affiliation{Condensed Matter Physics Center (IFIMAC) and Departamento de Física Teórica de la Materia Condensada,
Universidad Autónoma de Madrid, E-28049 Madrid, Spain}
\author{Jacob Linder}
\affiliation{Center for Quantum Spintronics, Department of Physics,\\ Norwegian University of Science and Technology, NO-7491 Trondheim, Norway}
\date{\today}



\begin{abstract}

When a spin-splitting field is introduced to a thin film superconductor, the spin currents polarized along the field couples to energy currents that can only decay via inelastic scattering. We study spin and energy injection into such a superconductor where spin-orbit impurity scattering yields inverse spin-Hall and spin-swapping currents. We show that the combined presence of a spin-splitting field, superconductivity, and inelastic scattering gives rise to a strong enhancement of the ordinary inverse spin-Hall effect, as well as unique inverse spin-Hall and spin-swapping signals orders of magnitude stronger than the ordinary inverse spin-Hall signal. These can be completely controlled by the orientation of the spin-splitting field, resulting in a long-range charge and spin accumulations detectable much further from the injector than in the normal-state. While the enhanced inverse spin-Hall signals offer a major improvement in spin detection sensitivity, the unique spin-swap signals can be utilized for designing devices where both the spin and current directions are controlled and altered throughout the geometry.

\end{abstract}


\maketitle



\prlsection{Introduction} Superconductors, while fascinating on their own, exhibit emergent quantum phenomena in combination with magnetic materials that are pursued for technological applications and fundamental interest \cite{bergeret_rmp_05, buzdin_rmp_05, eschrig_rpp_15, bergeret_rmp_18, holmes_roadmap_17, amundsen_arxiv_22}. This includes phenomena such as extreme sensitivity to electromagnetic fields \cite{jaklevic_prl_64} and heat \cite{kalenkov_prl_12, machon_prl_13, ozaeta_prl_14, kolenda_prl_16}, infinite magnetoresistance \cite{huertashernando_prl_02, li_prl_13}, qubits \cite{feofanov_nphys_10}, and dissipationless flow of spin \cite{keizer_nature_06, khaire_prl_10, robinson_science_10}.
Enhancing and measuring spin transport via superconductors are among the main aims of the field \cite{linder_nphys_15}.

While spin in normal-metals is carried by spin-polarized electrons, spin in superconductors can be carried either by the Cooper pair condensate \textit{or} by quasi-particle excitations \cite{linder_nphys_15}. Quasi-particle currents in superconductors resemble electron currents in normal-metals in that they are both dissipative, but differ qualitatively due to quasi-particles having a highly energy-dependent charge and velocity while their spin is constant. This feature causes spin transport via quasi-particles to depend strongly on whether decay occurs via spin-orbit scattering or magnetic impurities \cite{takahashi_prl_02, morten_prb_04, silaev_prl_15, wakamura_prl_14,johnsen_prb_21}.
When spin-polarizing a superconducting film by making its thickness substantially smaller than the penetration depth of a magnetic field \cite{anderson_pr_63, englert_prl_64, higgs_prl_64,meservey_physrep_94}, unique transport properties are revealed \cite{bergeret_rmp_18}, leading to, \textit{e.g.},~large and tunable thermoelectric effects \cite{machon_prl_13, ozaeta_prl_14,kolenda_prl_16, kolenda_prb_17, gonzalezruano_arxiv_23}. In such spin-split superconductors, quasi-particle spin currents couple to energy currents that are relaxed over much larger length scales via inelastic scattering \cite{bergeret_rmp_18,machon_prl_13,ozaeta_prl_14,quay_nphys_13}.

A key component in spin transport is the manner in which spin currents are detected. This is customarily done using the inverse spin-Hall effect \cite{dyakonov_pla_71, hirsch_prl_99} where a spin current is converted into a transverse electric voltage. The efficiency of the spin-to-charge conversion is quantified by a spin-Hall angle~$\theta_{\text{sH}}$. Previous works predicted superconductivity to cause slightly enhanced detection sensitivity \cite{espedal_prb_17}. Experiments have, on the other hand, observed an inverse spin-Hall effect that exceeds its normal-state value by three orders of magnitude \cite{wakamura_nmat_15, jeon_acsnano_20}. Owing to its intriguing effects on transport phenomena, introducing a spin-splitting field could have a profound effect on the spin-Hall effect and its inverse. However, this has not been investigated so far.

Here, we use Keldysh non-equilibrium Green's function theory \cite{serene_physrep_83, rammer_rmp_86, belzig_micro_99} to compute the inverse spin-Hall response of a spin-split superconductor (ssSC). Additionally, we compute the spin-swapping properties \cite{lifshits_prl_09} -- the conversion of a spin-polarized current flowing in one direction to a differently polarized spin current flowing in a perpendicular direction. We find a strong enhancement of the inverse spin-Hall signal, tunable via the orientation of the spin-splitting field. Moreover, unique types of inverse spin-Hall and spin-swapping signals appear in the ssSC, orders of magnitude stronger than those found in normal-metals and superconductors previously and measurable far away from the injector.
The control over both spin and current directions provided through the unique spin-swap effects offers flexibility in designing device geometries. This is useful for transporting spin signals through superconducting devices, and also for spin injection into other materials where non-equilibrium phenomena such as spin pumping \cite{tserkovnyak_rmp_05} and magnon currents \cite{han_nm_20,yuan_pr_22} can be studied. The large and tunable inverse spin-Hall signals provide the benefit of higher detection sensitivity, important due to the widespread use of spin currents in spintronics and related fields.

\begin{figure*}[t!]
    \centering
    \includegraphics[width=\textwidth]{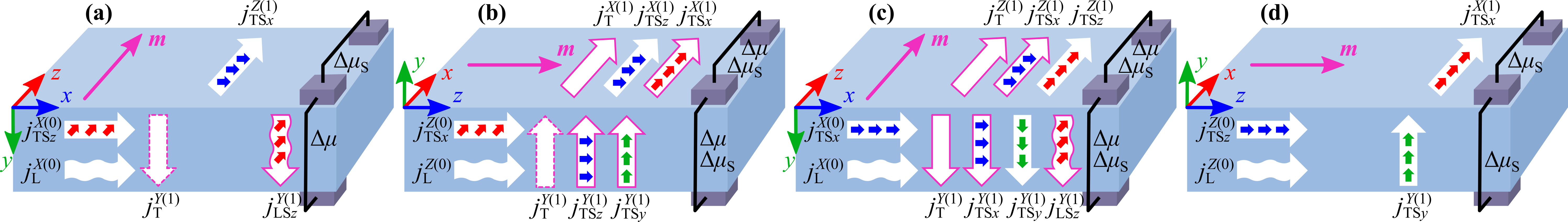}
    \caption{We inject a spin current $\ve{j}_{\text{TS}x_i}^{(0)}$ polarized along $x_i$ and an energy current $\ve{j}_{\text{L}}^{(0)}$ (white horizontal arrows) into a ssSC with an in-plane spin-splitting field $\ve{m}$ (pink arrow). In panels (a) and (b), the spin polarization (small arrows) of the injected spin current is perpendicular to the current direction, and in panels (c) and (d), it is parallel. From panel (a) to (b), and from panel (c) to (d), the spin-splitting field is rotated by $\pi/2$. We rotate the coordinate system so that $\ve{m}=m\ve{z}$ and the energy current always couples to the $z$ polarized spin current.
    The injected currents produce transversal currents through the inverse spin-Hall and spin-swap effects.
    The transversal currents that are present only in a ssSC are outlined by a pink solid line, and those that are renormalized by the spin-splitting field are outlined by a dashed pink line.
    The transversal charge and $z$ polarized spin-energy currents $\ve{j}_{\text{T}}^{(1)}$ and $\ve{j}_{\text{LS}z}^{(1)}$ produce charge accumulations $\Delta\mu$. The transversal spin currents $\ve{j}_{\text{TS}x_i}^{(1)}$ produce spin accumulations $\Delta\mu_{\text{S}}$. While transversal energy currents could in principle contribute to $\Delta\mu_{\text{S}}^z$, they are not present here.
    Note that although the white arrows point in the positive current direction, the directions of the currents can be positive or negative depending on the parameters.
    }
    \label{fig:model}
\end{figure*}

\prlsection{Theory}
We consider a ssSC connected to a normal-metal contact. When applying a spin-polarized voltage, both spin and energy quasi-particle currents are injected into the ssSC. The spin current decays inside the ssSC due to ordinary, spin-orbit, spin-flip, and inelastic scattering, while the energy current can only decay through inelastic scattering \cite{bergeret_rmp_18}. We assume the length of the ssSC to be larger than the inelastic scattering length, suppressing back-flow currents. The spin-orbit scattering also generates transversal currents through the inverse spin-Hall and spin-swap effects.

To study these transversal currents, we consider the Usadel equation
$\nabla_{\veR}\cdot\check{\ve{\mathcal{I}}}(\veR,\epsilon) = i[\check{\sigma}(\veR,\epsilon),\check{g}_{\text{av}}^{\text{s}}(\veR,\epsilon)]+\check{\mathcal{T}}(\veR,\epsilon)$
for the Keldysh space Green's function $\check{g}_{\text{av}}^{\text{s}}(\veR,\epsilon)$ including a matrix current \cite{bergeret_prb_16,huang_prb_18,Virtanen_PRB_2021}
\begin{align}
&\check{\ve{\mathcal{I}}}(\veR,\epsilon) =
     -D\Big(-\frac{i\kappa}{2}\left\{\hrhot\vehsigma\comtimes\check{g}_{\text{av}}^{\text{s}}(\veR,\epsilon)\nabla_{\veR}\check{g}_{\text{av}}^{\text{s}}(\veR,\epsilon)\right\} \notag\\
    &-\frac{\theta}{2}\left[\hrhot \vehsigma \comtimes\nabla_{\veR}\check{g}_{\text{av}}^{\text{s}}(\veR,\epsilon)\right]+\check{g}_{\text{av}}^{\text{s}}(\veR,\epsilon)\nabla_{\veR}\check{g}_{\text{av}}^{\text{s}}(\veR,\epsilon)\Big)
     \label{eq:current_operator}
\end{align}
with first order corrections in the spin-orbit parameter $\alpha$ entering through the normal-state spin-Hall and spin-swap angles $\theta$ and $\kappa$, respectively.
For its derivation, details about the calculation, physical observables, and the choice of parameters, see the Supplemental Material (SM) 
\footnote{See Supplemental Material for the derivation of the Usadel equation, kinetic equations, and the non-equilibrium charge and spin accumulations, and for results for the inverse spin-Hall signal at different spin-voltages. The Supplemental Material includes the additional references~\cite{Keizer_PRL_2006,Feshchenko_PhysRevAppl_2015,Strambini_NatCommun_2022}}.
Above, $D=\tau v_F^2 /3$ is the diffusion coefficient determined by the scattering time $\tau$ and the Fermi velocity $v_{\text{F}}$.
The torque $\check{\mathcal{T}}(\veR,\epsilon)$ arises from the first order corrections in the spin-orbit scattering, but only gives a nonzero contribution in the presence of supercurrents. We assume the retarded part of the Green's function to be constant in space, focusing only on the quasi-particle transport. 
The self-energy $\check{\sigma}(\veR,\epsilon)$ of a ssSC is given by $\hat{\sigma}_{\text{ssSC}}(\epsilon)=\epsilon\hrhot + \hat{\Delta}-\hat{\ve{\sigma}}\cdot\ve{m}$, where $\epsilon$ is the energy, $\hat{\Delta}=\text{diag}(\Delta,-\Delta,\Delta^* ,-\Delta^* )$ is the matrix introducing the superconducting gap $\Delta$, $\hat{\ve{\sigma}}=\text{diag}(\ve{\sigma},\ve{\sigma}^* )$, where $\ve{\sigma}$ is the vector of Pauli matrices, and $\ve{m}$ is the spin-splitting field. The superconducting gap is calculated self-consistently for the given $\ve{m}$. Additionally, we include spin-orbit, spin-flip, and inelastic scattering, respectively, through the self-energy terms
$\hat{\sigma}_{\text{so}}(\veR,\epsilon)  = \frac{i}{8\tau_{\text{so}}}\hrhot\vehsigma\cdot\check{g}_{\text{av}}^{\text{s}}(\veR,\epsilon)\hrhot\vehsigma,$
$\hat{\sigma}_{\text{sf}}(\veR,\epsilon) = \frac{i}{8\tau_{\text{sf}}}\vehsigma\cdot\check{g}_{\text{av}}^{\text{s}}(\veR,\epsilon)\vehsigma$, and
$\check{\sigma}_{\text{isct}}(\epsilon)=i\delta\text{diag}(\hrhot,-\hrhot)+2i\delta\tanh\left(\frac{\epsilon}{2T}\right)\text{antidiag}(\hrhot,0).$
Here, $\tau_{\text{so}}$ and $\tau_{\text{sf}}$ are the spin-orbit and spin-flip scattering times, $\delta$ determines the strength of the inelastic scattering, and $T$ is the temperature.
We have defined the matrices $\hat{\rho}_3 = \text{diag}(1,1,-1,-1)$, and $\hat{\rho}_0 = \text{diag}(1,1,1,1)$.

From the above, we derive the non-equilibrium charge and spin accumulations across the ssSC resulting from the transversal currents.
We choose to fix the spin-splitting field along $\ve{z}$. 
In this case, the energy current always couples to the $z$ polarized spin current, and the charge current to the $z$ polarized spin-energy current. 
The $x$ and $y$ polarized spin currents are also coupled together due to the precession of the spin around the spin-splitting field \cite{bergeret_rmp_18}.
In the following, we present the transversal currents $\ve{j}^{(1)}$ relevant for the charge and spin accumulations in terms of the injected currents $\ve{j}^{(0)}$. These are derived from the Keldysh part of the matrix current in Eq.~\eqref{eq:current_operator}, and result from first and zeroth order terms in the spin-orbit parameter $\alpha$, respectively.

\prlsection{The inverse spin-Hall effect}
In the inverse spin-Hall effect, a transversely polarized spin current is transformed into a transversal charge current resulting in a non-equilibrium charge accumulation across the superconductor \cite{dyakonov_pla_71, hirsch_prl_99}. In ssSCs, the charge accumulation can have an additional contribution from a spin-energy current polarized parallel to the spin-splitting field \cite{bergeret_rmp_18}. 
We first study how the inverse spin-Hall effect is renormalized in a ssSC compared to a superconductor (SC) and a normal-metal (NM), and then consider charge accumulations that only occur in the ssSC. 

\begin{figure}[t!]
    \centering
    \includegraphics[width=\columnwidth]{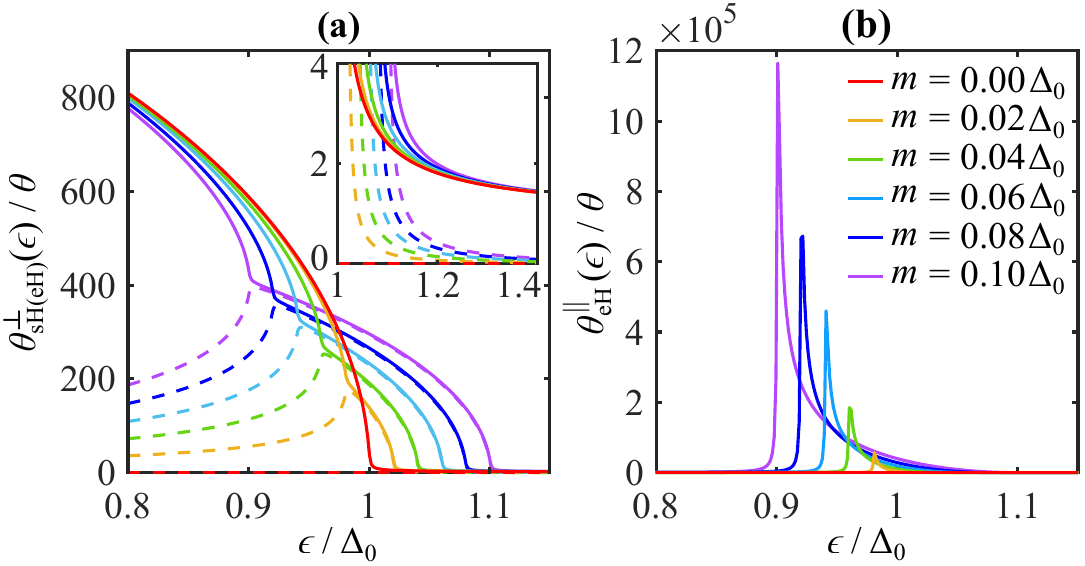}
    \caption{(a) The spin-Hall angle $\theta_{\text{sH}}^{\perp}$ (solid lines) and the energy-Hall angle $\theta_{\text{eH}}^{\perp}$ (dashed lines) for various spin-splitting fields $m$. Owing to inelastic scattering, there is a huge renormalization of $\theta_{\text{sH}}^{\perp}(\epsilon)$ and $\theta_{\text{eH}}^{\perp}(\epsilon)$ below the outer gap edge (rightmost kink). Above the outer gap, where both spin-up and spin-down quasi-particles are present, there is a weaker renormalization (inset). At large energies, $\theta_{\text{sH}}^{\perp}(\epsilon)\to\theta$ and $\theta_{\text{eH}}^{\perp}(\epsilon)\to0$.
    (b) The energy-Hall angle $\theta_{\text{eH}}^{\parallel}$ is strongly renormalized between the inner and outer gap edge. At large energies, $\theta_{\text{eH}}^{\parallel}(\epsilon)\to0$. 
    Note that $\theta_{\text{sH}}^{\perp}(\epsilon)$ is even in energy, while $\theta_{\text{eH}}^{\perp}(\epsilon)$ and $\theta_{\text{eH}}^{\parallel}(\epsilon)$ are odd in energy.
    The energy is normalized by the zero-temperature superconducting gap $\Delta_0$ at $m=0$, while the angles are normalized by the normal-state spin-Hall angle $\theta$. We consider zero temperature, so that $\Delta=\Delta_0$ at $m=0$.}
    \label{fig:angles}
\end{figure}

If the spin polarization of the injected current and the
spin-splitting field are both oriented along $\ve{z}$ and perpendicular to the direction of the injected current $\ve{x}$, the transversal charge current is out-of-plane (OOP) and given by
\begin{align}
    j_{\text{T}}^{Y(1)}(x,\epsilon) = -\theta_{\text{sH}}^{\perp}(\epsilon)j_{\text{TS}z}^{X(0)}(x,\epsilon)+\theta_{\text{eH}}^{\perp}(\epsilon)j_{\text{L}}^{X(0)}(x,\epsilon),
\end{align}
with the spin-Hall angle
\begin{align}
    \theta_{\text{sH}}^{\perp}(\epsilon) = 
    \begin{cases}
        \theta D\frac{N_{+}(\epsilon)D_{\text{L}}(\epsilon)- N_- (\epsilon) D_{\text{TS}z} (\epsilon)}{[D_{\text{L}}(\epsilon)]^2 - [D_{\text{TS}z}(\epsilon)]^2} \text{ for a ssSC,}\\
        \theta D N(\epsilon)/D_{\text{L}}(\epsilon)\text{ for a SC,}\\
        \theta \text{ for a NM,}
    \end{cases}
\end{align}
and the energy-Hall angle
\begin{align}
    \theta_{\text{eH}}^{\perp}(\epsilon) = 
    \begin{cases}
    \theta D\frac{N_{+}(\epsilon)D_{\text{TS}z} (\epsilon)- N_- (\epsilon)D_{\text{L}}(\epsilon)}{[D_{\text{L}}(\epsilon)]^2 - [D_{\text{TS}z} (\epsilon)]^2 } \text{ for a ssSC,}\\ 
    0 \text{ for a SC and a NM,}
    \end{cases}
\end{align}
both proportional to the spin-orbit parameter $\alpha$.
The OOP spin-energy current that also contributes to the charge accumulation is given by
\begin{align}
    j_{\text{LS}z}^{Y(1)}(x,\epsilon) = -\theta_{\text{sH}}^{\perp}(\epsilon)j_{\text{L}}^{X(0)}(x,\epsilon) + \theta_{\text{eH}}^{\perp}(\epsilon)j_{\text{TS}z}^{X(0)}(x,\epsilon).
\end{align}
These OOP currents are illustrated in Fig.~\ref{fig:model}(a).
Above, $N_+(\epsilon)$ and $N(\epsilon)$ are the density-of-states (DOS) normalized by their normal-state value in the ssSC and SC, respectively, $D_{\text{L}}(\epsilon)=D$ in the normal-state, and $N_-(\epsilon)$ and $D_{\text{TS}z}(\epsilon)$ are only non-zero in the presence of spin-splitting. Complete expressions are given in the SM.
The above spin-Hall and energy-Hall angles are plotted in Fig.~\ref{fig:angles}(a). 
For the given ratio between the inelastic scattering parameter and the zero-temperature superconducting gap, $\delta/\Delta_0=10^{-3}$, there is a two orders of magnitude increase in the spin-Hall and energy-Hall angles below the gap edge of the SC. There is also a large renormalization for energies between the inner and outer gap edges of the spin-split DOS. There is a smaller increase in the spin-Hall and energy-Hall angles above the outer gap where both spin-species are present. When increasing (decreasing) $\delta/\Delta_0$ by one order of magnitude, $\theta_{\text{sH}}^{\perp}$ and $\theta_{\text{eH}}^{\perp}$, approximately scaling as $(\delta/\Delta_0)^{-1}$, decrease (increase) by one order of magnitude (see SM).

If we rotate the magnetic field so that it is parallel to the propagation direction, the charge current is instead given by
\begin{align}
    j_{\text{T}}^{Y(1)}(z,\epsilon) = \theta_{\text{sH}}^x (\epsilon)j_{\text{TS}x}^{Z(0)}(z,\epsilon)-\theta_{\text{sH}}^y (\epsilon)j_{\text{TS}y}^{Z(0)}(z,\epsilon).
\end{align}
There is no spin-energy current contribution to the charge accumulation.
We have here defined the propagation direction of the injected current along $\ve{z}$ and its spin along $\ve{x}$. The precession of the spin around the spin-splitting field results in two spin-Hall angles for the current polarized along $\ve{x}$ and $\ve{y}$,
\begin{align}
    \theta_{\text{sH}}^x(\epsilon)& =
    \begin{cases}
    \theta D  \frac{N_+ (\epsilon)D_{\text{TS}x} (\epsilon)+N_-^{\text{I}}(\epsilon)D_{\text{TS}y} (\epsilon)}{[D_{\text{TS}x} (\epsilon)]^2 + [D_{\text{TS}y} (\epsilon)]^2 } \text{ for a ssSC,}\\
    \theta_{\text{sH}}(\epsilon) \text{ for a SC and NM,}
    \end{cases}\\
\theta_{\text{sH}}^y (\epsilon)&=
\begin{cases}
    \theta D  \frac{N_+ (\epsilon)D_{\text{TS}y} (\epsilon)-N_-^{\text{I}}(\epsilon)D_{\text{TS}x} (\epsilon)}{[D_{\text{TS}x} (\epsilon)]^2 + [D_{\text{TS}y} (\epsilon)]^2 } \text{ for a ssSC,}\\
    0 \text{ for a SC and a NM,}
\end{cases}
\end{align}
both proportional to $\alpha$.
Above, $D_{\text{TS}x}(\epsilon)=D_{\text{TS}z}(\epsilon)$ in the absence of spin-splitting, while $N_-^{\text{I}}(\epsilon)$ and $D_{\text{TS}y}(\epsilon)$ are only non-zero in the presence of spin-splitting.
The OOP charge current is illustrated in Fig.~\ref{fig:model}(b). 
When increasing the spin-splitting field, the above spin-Hall angles are suppressed compared to $\theta_{\text{sH}}^{\perp}(\epsilon)$.
The charge accumulation can therefore be controlled by rotating the spin-splitting field between the configurations in Figs.~\ref{fig:model}(a) and~(b). 

We next consider charge accumulations that only appear in the presence of spin-splitting. 
Consider the case where the spin of the injected current is parallel to its propagation direction $\ve{x}$, and the spin-splitting field is perpendicular to these. In this case, we find OOP charge and spin-energy currents
\begin{align}
    j_{\text{T}}^{Y(1)}(x,\epsilon) &= -\theta_{\text{eH}}^{\parallel}j_{\text{L}}^{X(0)}(x,\epsilon),\\
    j_{\text{LS}z}^{Y(1)}(x,\epsilon) &= -[N_+(\epsilon)/N_-(\epsilon)]\theta_{\text{eH}}^{\parallel}j_{\text{L}}^{X(0)}(x,\epsilon)
\end{align}
that only have a contribution from the injected energy current. The energy-Hall angle is given by
\begin{align}
    \theta_{\text{eH}}^{\parallel} &=
    \begin{cases}
        \theta D\frac{N_{-}(\epsilon)D_{\text{L}}(\epsilon)}{[D_{\text{L}}(\epsilon)]^2-[D_{\text{TS}z}(\epsilon)]^2}, \text{ for a ssSC,}\\
        0 \text{ for a SC and a NM,}
    \end{cases}
    \label{eq:theta_eH_parallel}
\end{align}
and is proportional to $\alpha$.
While the spin-energy current is finite also in the absence of spin-splitting due to $N_+(\epsilon)$ being finite, it only gives a contribution to the charge accumulation in the presence of spin-splitting.
The above OOP charge and spin-energy currents are illustrated in Fig.~\ref{fig:model}(c).
They disappear when rotating the spin-splitting field to the parallel orientation, as shown in Fig.~\ref{fig:model}(d).
The corresponding energy-Hall angle is plotted in Fig.~\ref{fig:angles}(b). Owing to the inelastic scattering, there is a huge renormalization between the inner and outer gap edges. The angle of the spin-energy current is similarly renormalized between the inner and outer gap, but is instead odd in energy. When increasing (decreasing) $\delta/\Delta_0$ by one order of magnitude, $\theta_{\text{eH}}^{\parallel}$, approximately scaling as $(\delta/\Delta_0)^{-2}$, decreases (increases) by two orders of magnitude.

Additionally, in-plane (IP) charge currents 
\begin{align}
    j_{\text{T}}^{X(1)}(z,\epsilon) &=-\theta_{\text{sH}}^x(\epsilon)j_{\text{TS}y}^{Z(0)}(z,\epsilon)-\theta_{\text{sH}}^y(\epsilon)j_{\text{TS}x}^{Z(0)}(z,\epsilon),\\
    j_{\text{T}}^{Z(1)}(x,\epsilon) &=\theta_{\text{sH}}^x(\epsilon)j_{\text{TS}y}^{X(0)}(X,\epsilon)+\theta_{\text{sH}}^y(\epsilon)j_{\text{TS}x}^{X(0)}(x,\epsilon),
\end{align}
exist when the spin-splitting field is perpendicular to the injected spin, regardless of their orientation with respect to the direction of the injected current. This is shown in Figs.~\ref{fig:model}(b) and~(c), respectively. These currents disappear when rotating the spin-splitting field by $\pi/2$, see Figs.~\ref{fig:model}(a) and~(d).
Thus, several transversal currents appear that are only present in a ssSC and can be controlled by the orientation of the spin-splitting field.

\begin{figure}[t]
    \centering
    \includegraphics[width=\columnwidth]{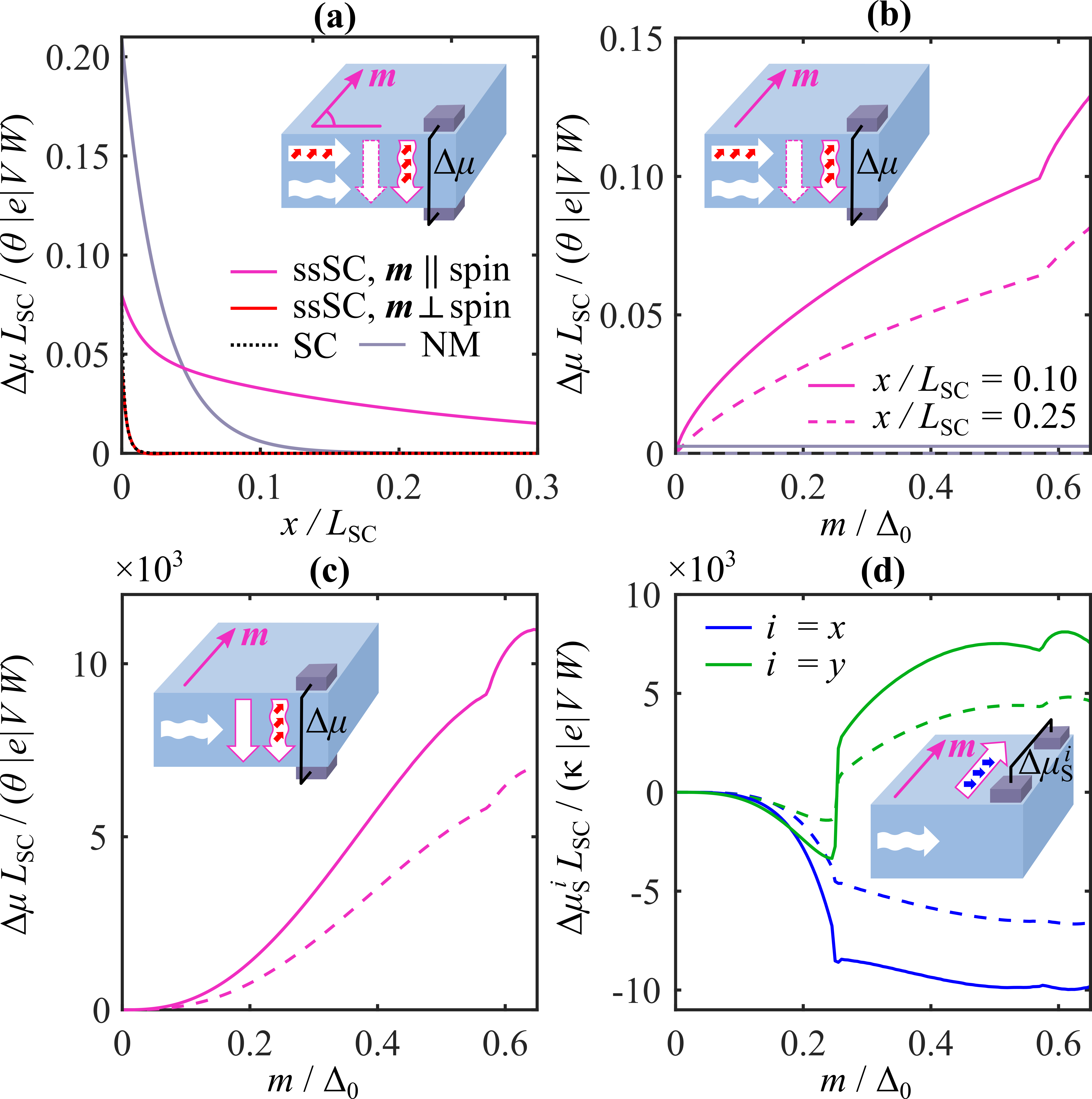}
    \caption{(a) When the injected spin is oriented along the spin-splitting field ($\ve{m}\parallel$~spin, Fig.~\ref{fig:model}(a)), the OOP charge accumulation $\Delta\mu$ can be detected for much longer distances inside a ssSC than inside a NM, and (b) increases with increasing spin-splitting field. When the spin-splitting field is rotated by $\pi/2$ ($\ve{m}\perp$~spin, Fig.~\ref{fig:model}(b)), the charge accumulation is strongly suppressed and behaves as in the absence of spin-splitting. The charge accumulation is normalized by $\theta|e|V W /L_{\text{SC}}$, where $e$ is the electron charge, $V=(V_{\uparrow}-V_{\downarrow})$ is the spin-voltage in the injector, $W$ is the distance between the detectors, and $L_{\text{SC}}$ is the length of the ssSC. 
    (c)~A charge accumulation five orders of magnitude larger than in panel~(b) can be obtained in the configuration in Fig.~\ref{fig:model}(c) due to the huge renormalization of the spin-Hall angle shown in Fig.~\ref{fig:angles}(b).
    (d) A spin accumulation $\Delta\mu_{\text{S}}^{x(y)}$ of the same order of magnitude results from the IP spin current $j_{\text{TS}x}^{Z(1)}$ in Fig.~\ref{fig:model}(c). Spin accumulations in Fig.~\ref{fig:model}(b) resulting from $j_{\text{TS}x}^{X(1)}\big(j_{\text{TS}y}^{Y(1)}\big)$ are obtained by letting $\Delta\mu_{\text{S}}^{x},\Delta\mu_{\text{S}}^{y}\to-\Delta\mu_{\text{S}}^{x},-\Delta\mu_{\text{S}}^{y}\; (-\Delta\mu_{\text{S}}^{y},\Delta\mu_{\text{S}}^{x})$. The schematics only include the incoming and transversal currents that contribute to the corresponding charge and spin accumulations. For panels~(b)-(d), solid and dotted curves refer to positions $x/L_{\text{SC}}=0.10$ and $x/L_{\text{SC}}=0.25$, respectively. We consider $m=0.1\Delta_0$, where $\Delta_0$ is the zero-temperature gap at $m=0$, $|e|V=2.5\Delta_0$, and $T = T_c/4$. }
    \label{fig:accumulations}
\end{figure}

\prlsection{Spin-swapping}
A spin-swap current is a transversal spin current that appears due to an injected spin current \cite{lifshits_prl_09}. As a result, there is a non-equilibrium spin accumulation across the ssSC.
We first consider transversal currents where the propagation direction and spin polarization are perpendicular to each other. 
If we inject a spin current with spin polarization and propagation direction along the $x_i$ and $x_j$ axes, respectively, we produce an IP spin-swap current
$j_{\text{TS}x_j}^{X_i(1)}(X_j,\epsilon) = -\kappa j_{\text{TS}x_i}^{X_j(0)}(X_j ,\epsilon)$ where the indices $i$ and $j$ are swapped compared to the incoming current. This is illustrated in Figs.~\ref{fig:model}(a) and~(b).
When the spin-splitting field is perpendicular to the spin-polarization of the injected current, an additional OOP spin-swap current appears, as shown in Figs.~\ref{fig:model}(b) and~(c). These still follow the same expression as above, but are absent when rotating the spin-splitting field by $\pi/2$ to the configuration in Figs.~\ref{fig:model}(a) and~(d), respectively.
In the case when the spin polarization of the injected current is along the current direction, but the spin-splitting field is perpendicular to these, we find an IP spin current $j_{\text{TS}x}^{Z(1)}(x,\epsilon) = -\kappa_{\text{es}}j_{\text{L}}^{X(0)}(x,\epsilon)$ with a spin-swap angle
\begin{align}
    \kappa_{\text{es}}&=
    \begin{cases}
        \kappa \frac{D_{\text{L}}(\epsilon)D_{\text{TS}}^z(\epsilon)}{[D_{\text{L}}(\epsilon)]^2-[D_{\text{TS}}^z(\epsilon)]^2} \text{ for a ssSC,}\\
        0 \text{ for a SC and a NM,}
    \end{cases}
    \label{eq:kappa_es}
\end{align}
proportional to $\alpha$, where only the energy current contributes. This IP current is shown in Fig.~\ref{fig:model}(c), and disappears when the spin-splitting field is rotated to the configuration in Fig.~\ref{fig:model}(d). The above spin-swap angle is only non-zero below the outer gap edge, is greatly renormalized between the inner and outer gap edges, and scales as $(\delta/\Delta_0)^{-2}$, similar to the energy-Hall angle in Fig.~\ref{fig:angles}(b).

We next consider the transversal currents that carry spin polarized along their propagation direction.
As a result of an incoming spin current polarized along its propagation direction, we find IP and OOP transversal spin currents $j_{\text{TS}x_i}^{X_i(1)}(X_j,\epsilon) = \kappa j_{\text{TS}x_j}^{X_j(0)}(X_j ,\epsilon)$ that only depend on the injected spin current via the normal-state spin-swap angle, see Fig.~\ref{fig:model}(c) and~(d). 
However, similar currents also appear when the spin polarization of the injected current is perpendicular to its propagation direction if the spin-splitting field is parallel to the incoming current. In this case, only the energy current contributes to the transversal spin currents $j_{\text{TS}x_i}^{X_i(1)}(z,\epsilon) = \kappa_{\text{es}}j_{\text{L}}^{Z(0)}(z,\epsilon)$
through a strongly renormalized spin-swap coefficient.
These currents are illustrated in Fig.~\ref{fig:model}(b), and disappear in Fig.~\ref{fig:model}(a) where the spin-splitting field is rotated.
Although the ordinary spin-swap angles are unaffected by spin-splitting and superconductivity, additional transversal currents appear that depend either on the normal-state spin-swap angle or a strongly renormalized one.

\prlsection{Charge and spin accumulations}
We next study the resulting non-equilibrium charge and spin accumulations measured across the transversal IP and OOP directions.
We focus cases where spin-splitting couples the transversal currents to the injected energy current. 
Since the inelastic scattering rate is typically much slower than the spin-orbit and spin-flip scattering rates, the energy current survives far into the ssSC compared to the injected spin current \cite{bergeret_rmp_18}.
In Figs.~\ref{fig:accumulations}(a) and~(b), we show how the contribution from the long-range energy current gives rise to an inverse spin-Hall signal that survives far inside the ssSC. The signal remains orders of magnitude larger than in the normal-state even when the spin voltage $|e|V$ is lowered toward the gap $\Delta_0$ (see SM). 
Without coupling between the injected spin and energy currents, the charge accumulation is small because spin injection is forbidden below the superconducting gap and decays rapidly because the quasi-particle spin currents are sensitive to spin-flip scattering \cite{morten_prb_04,johnsen_prb_21}. Similar to our predictions, large inverse spin-Hall signals have recently been observed experimentally in ssSCs \cite{jeon_acsnano_20}.

As shown in Figs.~\ref{fig:accumulations}(c) and~(d), charge and spin accumulations orders of magnitude larger than the ordinary inverse spin-Hall signal can be obtained as a result of the strongly renormalized spin-Hall and spin-swap angles in Eqs.~\eqref{eq:theta_eH_parallel} and~\eqref{eq:kappa_es}. Owing to inelastic scattering, these are massively renormalized between the inner and outer gap edges in the spin-split DOS, as demonstrated in Fig.~\ref{fig:angles}(b). The enhancement happens when the spin-splitting field is oriented perpendicular to the spin of the injected spin current so that only the energy current contributes to the detected signal. Especially intriguing is the possibility of generating large OOP accumulations of OOP spins from pure energy currents (see $j_{\text{TS}y}^{Y(1)}$ in Fig.~\ref{fig:model}(b)). This allows injection of OOP spins into an adjacent material, \textit{e.g.}, in a stack without placing the superconductor in proximity to magnet with perpendicular magnetization, thereby reducing additional stray fields besides those stemming from the generated spins.

\prlsection{Concluding remarks}
In addition to enhancing the inverse spin-Hall signal, we find that a spin-splitting field leads to unique inverse spin-Hall and spin-swap signals, orders of magnitude larger than the ordinary inverse spin-Hall signal.
These results offer major improvements in spin detection sensitivity and opportunities for designing new device geometries where both spin and current directions can be controlled via the orientation of the spin-splitting field.

 \begin{acknowledgments}
\prlsection{Acknowledgments}
We thank J. Tjernshaugen and M. Amundsen for useful discussions. L.J.K. and J.L. acknowledge financial support from the Research
Council of Norway through Grant No. 323766 and its Centres
of Excellence funding scheme Grant No. 262633 “QuSpin.” Support from
Sigma2 - the National Infrastructure for High Performance
Computing and Data Storage in Norway, project NN9577K, is acknowledged.
L.J.K. acknowledges financial support from the Spanish Ministry for Science and Innovation—AEI Grant No. CEX2018-000805-M (through the “Maria de Maeztu” Programme for Units of Excellence in R\&D) and Grant No. RYC2021-031063-I funded by MCIN/AEI and “European Union Next Generation EU/PRTR”.
 \end{acknowledgments}




%


\onecolumngrid
\newpage

\begin{center}
\noindent
{\large\textbf{Supplemental Material to: Inverse spin-Hall effect and spin-swapping in spin-split superconductors}}\\
\vspace{1em}Lina Johnsen Kamra$^{1,2,*}$ and Jacob Linder$^1$\\
\vspace{0.5em}{\small$^1$\textit{Center for Quantum Spintronics, Department of Physics,}\\ \textit{Norwegian
University of Science and Technology, NO-7491 Trondheim, Norway}\\
$^2$\textit{Condensed Matter Physics Center (IFIMAC) and Departamento de F\'{i}sica Te\'{o}rica de la Materia Condensada,}\\
\textit{Universidad Aut\'{o}noma de Madrid, E-28049 Madrid, Spain}\\
(Dated: \today)}
\end{center}
\vspace{1.5em}

\noindent
We here outline the derivation of the Usadel equation with corrections to the first order in the spin-orbit parameter $\alpha$ (Sec. \hyperlink{sec:Usadel}{I}), provide details about the numerical solution of the kinetic equations (Sec.~\hyperlink{sec:kinetic}{II}), give expressions for the non-equilibrium charge and spin accumulations (Sec.~\hyperlink{sec:non-equilibrium_accumulations}{III}), and provide results for the inverse spin-Hall signal at different spin-voltages (Sec.~\hyperlink{sec:voltages}{IV}).

\section{I.\:\:The Usadel equation}

\hypertarget{sec:Usadel}{Our} starting point for deriving the Usadel equation given in Eq.~(1) in the main text is the continuum Hamiltonian
\begin{align}
    H(\ver,t)&=\int d\ver\: \sum_{\sigma}\psi_{\sigma}^{\dagger} (\ver,t)\Big(-\frac{1}{2m}\nabla_{\ver}^2 -\mu\Big)\psi_{\sigma}(\ver,t)
    +\frac{1}{2}\int d\ver \: \big[\Delta(\ver)\psi^{\dagger}_{\uparrow}(\ver,t)\psi_{\downarrow}^{\dagger}(\ver,t)+\text{h.c.}\big]\notag\\
    &+\int d\ver\: \sum_{\sigma,\sigma'}\psi_{\sigma}^{\dagger}(\ver,t)[\ve{m}(\ver)\cdot\ve{\sigma}]_{\sigma,\sigma'}\psi(\ver,t)
    +\int d\ver\: \sum_{\sigma,\sigma'}\psi_{\sigma}^{\dagger} (\ver,t)U_{\sigma,\sigma'}^{\text{tot}}(\ver)\psi_{\sigma'}(\ver,t),
\label{eq:hamiltonian}
\end{align}
where $\psi_{\sigma}^{(\dagger)}(\ver,t)$ is a field operator annihilating (creating) a spin-$\sigma$ electron at position $\ver$ and time $t$. 
The first term includes the kinetic energy for electrons of mass $m$, and the chemical potential $\mu$.
The second term introduces superconductivity, where $\Delta(\ver)= V\left<\psi_{\uparrow}(\ver)\psi_{\downarrow}(\ver)\right>$ is the mean-field superconducting gap. 
The third term introduces a spin-splitting field $\ve{m}(\ver)$.
The last term introduces the total scattering potential from the impurities $U_{\sigma,\sigma'}^{\text{tot}}(\ve{r})$. Above, $\ve{\sigma}$ is the vector of Pauli matrices.

We define the retarded, advanced and Keldysh Green's functions in Nambu~$\otimes$~spin space as
\begin{align}
    [\hat{G}^R (1,2)]_{i,j}
    =&-i\Theta(t_1 -t_2 )
    \sum_k (\hrhot)_{ik}\big<\big\{[{\psi}(1)]_k , [\hat{\psi}^{\dagger} (2)]_j \big\}\big>,~\label{eq:def_GR}\\
    [\hat{G}^A (1,2)]_{i,j} 
    =&\phantom{+}i\Theta(t_2 -t_1 )
    \sum_k (\hrhot)_{ik}\big<\big\{[\hat{\psi}(1)]_k , [\hat{\psi}^{\dagger} (2)]_j \big\}\big>,\label{eq:def_GA}\\
    [\hat{G}^K (1,2)]_{i,j} 
    =&-i\sum_k (\hrhot)_{ik}\big<\big[[\hat{\psi}(1)]_k , [\hat{\psi}^{\dagger} (2)]_j \big]\big>,\label{eq:def_GK}
\end{align}
respectively,
where $(1,2)$ is short-hand notation for $(\ver_1 ,t_1 ,\ver_2 ,t_2 )$, $\hat{\rho}_3 =\text{diag}(1,1,-1,-1)$, and we have defined a basis
\be
    \hat{\psi}(\ver,t)=[\psi_{\uparrow}(\ver,t)\hspace{0.5em}\psi_{\downarrow}(\ver,t)\hspace{0.5em}\psi^{\dagger}_{\uparrow}(\ver,t)\hspace{0.5em}\psi^{\dagger}_{\downarrow}(\ver,t)]^T .
\ee 
The above Green's functions are elements of the Keldysh space Green's function
\begin{align}
  \check{G}(1,2) =
  \begin{pmatrix}
      \hat{G}^{\text{R}}(1,2) & \hat{G}^{\text{K}}(1,2) \\
      0 & \hat{G}^{\text{A}}(1,2)
  \end{pmatrix}.
\end{align} 
Before we start introducing higher order corrections, our approach follows the one described in Ref.~\cite{johnsen_prb_21}.
From the Heisenberg equations of motion for the field operators, we find the equations of motion for the Keldysh space Green's function
\begin{align}
    [i\partial_{t_1}\hrhot -\hat{H}(\ver_1)]\check{G}(1,2)&=\delta(1-2)\check{\rho}_0,\label{eq:kinetic_eq_1}\\
    \check{G}(1,2)[i\partial_{t_2}\hrhot -\hrhot\hat{H}(\ver_2 )\hrhot]^{\dagger}&=\delta(1-2)\check{\rho}_0 .\label{eq:kinetic_eq_2}
\end{align}
where
\begin{align}
    \hat{H}(\ver)&=\left(-\frac{1}{2m}\nabla_{\ver}^2-\mu\right)\hrhoz
    -\hat{\Delta}(\ver)+ \hat{\ve{\sigma}}\cdot \ve{m}(\ver)+\hat{U}_{\text{tot}}(\ver)
     .
\end{align}
Above, $\hat{\rho}_0$ is the $4\times4$ unit matrix, and $\hat{\ve{\sigma}}=\text{diag}(\ve{\sigma},\ve{\sigma}^{*})$. The scattering potential matrix $\hat{U}_{\text{tot}}(\ver)=U(\ver)+\hat{U}_{\text{so}}(\ver)+\hat{U}_{\text{sf}}(\ver)$ describe ordinary scattering on non-magnetic impurities, spin-orbit scattering on non-magnetic impurities, and spin-flip scattering on magnetic impurities, respectively.
The scattering potentials are given by
\begin{align}
U(\ve{r})&=\sum_i u(\ve{r}-\ve{r}_i ),\\
\hat{U}_{\text{so}}(\ver)&=\sum_i i\alpha [\hat{\rho}_3 \hat{\ve{\sigma}}\times\nabla_{\ve{r}}u(\ve{r}-\ve{r}_i )]\cdot\nabla_{\ve{r}},\\
\hat{U}_{\text{sf}}(\ver)&=\sum_i u_{\text{m}} (\ver-\ver_i )\vehsigma\cdot \ve{S}_i,
\end{align}
where $u(\ve{r}-\ve{r}_i )$ and $u_{\text{m}} (\ve{r}-\ve{r}_i )$ are the scattering potentials of a single non-magnetic and magnetic impurity  at position $\ve{r}_i$, respectively, and $\ve{S}_i$ is the spin of the magnetic impurity.

In order to solve Eqs.~\eqref{eq:kinetic_eq_1} and~\eqref{eq:kinetic_eq_2}, we must replace the impurity potentials by self-energies. To do this, we split the Hamiltonian into two parts, $\hat{H}(\ver)=\hat{H}_0 (\ver) +\hat{U}_{\text{tot}}(\ver)$, where $\hat{H}_0 (\ver)$ describes the system in the absence of impurity scattering. The self-energies are introduced through the Dyson equations
\begin{align}
\check{G}(1,2)&=\check{G}_0 (1,2)+\check{G}_0\bullet\hat{\Sigma}\bullet\check{G}(1,2),\\
\check{G}(1,2)&=\check{G}_0 (1,2)+\check{G}\bullet\hat{\Sigma}^{\dagger }\bullet\check{G}_0 (1,2),
\end{align}
where the self-energies are defined as $\hat{\Sigma}(1,2)=\delta(1-2)\hat{U}_{\text{tot}}(\ver_2 )$.
Above, $\check{G}_0 (1,2)$ is the Green's function in the absence of impurity scattering, and we have introduced the bullet product
\be
A\bullet B(1,2)=\int d3\: A(1,3)B(3,2).
\ee
We solve the Dyson equations iteratively beyond the self-consistent Born approximation up to order $\mathcal{O}[(\hat{\Sigma}\bullet\check{G})^3]$ and $\mathcal{O}[(\check{G}\bullet\hat{\Sigma}^{\dagger})^3]$ following a similar approach as Ref.~\cite{espedal_prb_17,bergeret_prb_16,huang_prb_18}. Since we are not interested in one specific impurity configuration, we take the average over all impurities, 
\begin{align}
\Big<\hdots\Big>_{\text{av}}=\prod_{n=1}^N \left(\frac{1}{\mathcal{V}}\int d\ver_n \: \right)\hdots,
\end{align}
where $\mathcal{V}$ is the volume of the system. We assume that the Green's function is approximately equal to its impurity-averaged value.
By acting with $[i\partial_{t_1}\hrhot -\hat{H}_0 (\ver_1 )]$ and $[i\partial_{t_2}\hrhot -\hrhot\hat{H}_0 (\ver_2 )\hrhot]$ on the resulting equations, we obtain expressions similar to Eqs.~\eqref{eq:kinetic_eq_1} and~\eqref{eq:kinetic_eq_2}, where the impurity potentials are replaced by expressions involving self-energies and impurity averaged Green's functions $\check{G}_{\text{av}}(1,2)$. Subtracting the two equations, we find that
\begin{align}
    &[i\partial_{t_1}\hrhot -\hat{H}_0 (\ver_1 )]\check{G}_{\text{av}}(1,2)
    -\check{G}_{\text{av}}(1,2)[i\partial_{t_2}\hrhot -\hrhot\hat{H}_0 (\ver_2 )\hrhot]^{\dagger}\notag\\
   &-[\langle\hat{\Sigma}\bullet\check{G}_{\text{av}}\bullet\hat{\Sigma}\rangle_{\text{av}}\combullet\check{G}_{\text{av}} ](1,2)
   -[\langle\hat{\Sigma}\bullet\check{G}_{\text{av}}\bullet\hat{\Sigma}\bullet\check{G}_{\text{av}}\bullet\hat{\Sigma}\rangle_{\text{av}}\combullet\check{G}_{\text{av}} ](1,2)=0.\label{eq:equations_averaged}
\end{align}
We will now make a series of approximations to this equation.

We first introduce center-of-mass and relative coordinates $\veR=(\ver_1 +\ver_2 )/2$ and $\ver=\ver_1 -\ver_2$, and absolute and relative time coordinates $T=(t_1 +t_2 )/2$ and $t=t_1 - t_2 $. The Green's function is assumed to be independent of the absolute time coordinate.
We introduce the Fourier transform and its inverse, 
\begin{align}
\check{G}_{\text{av}}(\veR,\ve{p},\epsilon)&=\int d\ve{r}\int dt \: e^{-i\ve{p}\cdot\ve{r}+i\epsilon t}\check{G}_{\text{av}}(\veR,\ve{r},t) ,\label{eq:FT}\\
\check{G}_{\text{av}}(\veR,\ve{r},t)&=\int \frac{d\ve{p}}{(2\pi)^3}\int \frac{d\epsilon}{2\pi} e^{i\ve{p}\cdot\ve{r}-i\epsilon t} \check{G}_{\text{av}}(\veR,\ve{p},\epsilon) .\label{eq:iFT}
\end{align}
Assuming all quantities to vary slowly compared to the Fermi wavelength, the Fourier transform of the bullet product between two functions $A(\veR,\vep,\epsilon)$ and $B(\veR,\vep,\epsilon)$ can be expressed through the first order gradient approximation
\begin{align}
A\bullet B(\veR,\vep,\epsilon)=&A(\veR,\vep,\epsilon) B(\veR,\vep,\epsilon) +\frac{i}{2}[\nabla_{\veR}A(\veR,\vep,\epsilon)\cdot\nabla_{\vep}B(\veR,\vep,\epsilon)
-\nabla_{\vep}A(\veR,\vep,\epsilon)\cdot\nabla_{\veR}B(\veR,\vep,\epsilon)].
\label{eq:gradient_approximation}
\end{align}
Next, we assume that the absolute value of the momentum $\ve{p}$ is approximately equal to the Fermi momentum $p_{\text{F}}$, so that we can apply the quasi-classical approximation
\be
\int\frac{d\ve{p}}{(2\pi)^3}\:\check{G}_{\text{av}}(\veR,\ve{p},\epsilon)\approx N_0 \int d\xi_{p_{\text{F}}} \:\int \frac{d\ve{e}_{p_{\text{F}}}}{4\pi}\:\check{G}_{\text{av}}(\veR, \ve{p}_{\text{F}},\epsilon).\label{eq:quasi-cl_approx}
\ee
Above, $N_0$ is the DOS at the Fermi level, $\xi_{p_{\text{F}}}=p_{\text{F}}^2/2m$, and $\ve{e}_{p_{\text{F}}}=\ve{p}_{\text{F}}/p_{\text{F}}$ describes the direction of the momentum.
We use the short-hand notation
$\left<\hdots\right>_{p_{\text{F}}}=\int (d\ve{e}_{p_{\text{F}}}/4\pi)\:$ for the average over all directions of the momentum.
We moreover introduce the quasi-classical Green's function
\be
\check{g}_{\text{av}}(\ve{R},\ve{p}_{\text{F}} ,\epsilon )= \frac{i}{\pi}\int d\xi_{p_{\text{F}}} \: \check{G}_{\text{av}}(\ve{R},\ve{p}_{\text{F}},\epsilon).\label{eq:quasi-cl_GF}
\ee
In the diffusive limit, the quasi-classical Green's function can be approximated as
\be
\check{g}_{\text{av}}(\ve{R},\ve{p}_{\text{F}} ,\epsilon)\approx\check{g}_{\text{av}}^{\text{s}} (\ve{R},\epsilon)+\ve{e}_{p_{\text{F}}}\cdot\check{\ve{g}}_{\text{av}}^{\text{p}} (\ve{R},\epsilon).\label{eq:g->gs+gp}
\ee
After applying all these approximations to Eq.~\eqref{eq:equations_averaged}, we separate out the even contributions in $\vece_{p_{\text{F}}}$ by averaging over all $\vece_{p_{\text{F}}}$. We find that
\begin{align}
    &[\epsilon\hrhot+\hat{\Delta}(\veR)-\hat{\ve{\sigma}}\cdot\ve{m},\check{g}^{\text{s}}_{\text{av}}(\veR,\epsilon)]+\frac{iv_{\text{F}}}{3}\nabla_{\veR}\cdot \check{\ve{g}}_{\text{av}}^{\text{p}}(\veR,\epsilon)\notag\\ 
    &-\int\frac{d\ve{e}_{p_{\text{F}}}}{4\pi}\:[\check{\sigma}_{\text{av}}^{\text{tot}}(\veR,\ve{p}_{\text{F}},\epsilon),\check{g}_{\text{av}}(\veR,\ve{p}_{\text{F}},\epsilon)]
    +\frac{i}{2}\int\frac{d\ve{e}_{p_{\text{F}}}}{4\pi}\:\nabla_{\veR}\cdot\{\nabla_{\ve{p}}\check{\sigma}_{\text{av}}^{\text{tot}}(\veR,\ve{p},\epsilon)\big|_{\ve{p}=\ve{p}_{\text{F}}},\check{g}_{\text{av}}(\veR,\ve{p}_{\text{F}},\epsilon)\}
    = 0,\label{eq:even}
\end{align}
where $v_{\text{F}}=p_{\text{F}}/m$ is the Fermi velocity.
We next separate out the odd contributions in $\vece_{p_{\text{F}}}$ by multiplying the equation by $\vece_{p_{\text{F}}}$ before the averaging, which gives
\begin{align}
    &\frac{1}{3}[\epsilon\hrhot+\hat{\Delta}(\veR)-\hat{\ve{\sigma}}\cdot\ve{m},\check{\ve{g}}^{\text{p}}_{\text{av}}(\veR,\epsilon)]+\frac{iv_{\text{F}}}{3}\nabla_{\veR}\cdot\nabla_{\veR} \check{g}_{\text{av}}^{\text{s}}(\veR,\epsilon)
    -\int\frac{d\ve{e}_{p_{\text{F}}}}{4\pi}\:\vece_{p_{\text{F}}}[\check{\sigma}_{\text{av}}^{\text{tot}}(\veR,\ve{p}_{\text{F}},\epsilon),\check{g}_{\text{av}}(\veR,\ve{p}_{\text{F}},\epsilon)] 
    = 0.\label{eq:odd}
\end{align}
We have included terms to the zeroth order in the gradient approximation in the odd equation and to first order in the even equation. The reason will become clear later on. 

The most tricky part of solving Eqs.~\eqref{eq:even} and~\eqref{eq:odd} is to evaluate the self-energies
\begin{align}
    \check{\sigma}_{\text{av}}^{\text{tot}}(\veR,\ve{p},\epsilon) = \langle\check{\Sigma}\bullet\check{G}_{\text{av}}\bullet\check{\Sigma}\rangle_{\text{av}}(\veR,\ve{p},\epsilon)+ \langle\check{\Sigma}\bullet\check{G}_{\text{av}}\bullet\check{\Sigma}\bullet\check{G}_{\text{av}}\bullet\check{\Sigma}\rangle_{\text{av}}(\veR,\ve{p},\epsilon).
\end{align}
Our starting point is is the real space expression for the self-energies 
\begin{align}
    \check{\sigma}_{\text{av}}^{\text{tot}}(1,2)&=
    \left(\prod_i \frac{1}{\nu}\int d\ver_i \right)\hat{U}^{\text{tot}}(\ver_1 )\check{G}_{\text{av}}(1,2)\hat{U}^{\text{tot}}(\ver_2 )\notag\\
    &+\left(\prod_i \frac{1}{\nu}\int d\ver_i \right)\int d3\:\hat{U}^{\text{tot}}(\ver_1 )\check{G}_{\text{av}}(1,3)\hat{U}^{\text{tot}}(\ver_3 )\check{G}_{\text{av}}(3,2)\hat{U}^{\text{tot}}(\ver_2 )
\end{align}
We neglect scatterings including more than one impurity (\textit{i.e.} all of the impurity potentials in a single term has the same impurity index $i$). 
By following the same steps as described above for arriving at the even and odd equations, we find the following expressions for the self-energies: 
\newline
1) For ordinary scattering to the second order in the impurity potential, the self-energy is
\begin{align}
    \check{\sigma}^{u^2}(\veR,\ve{p},\epsilon) = -\frac{i}{2}\left<\frac{1}{\tau(\ve{p}-\ve{q}_{\text{F}})}\right>_{q_{\text{F}}}\check{\ve{g}}_{\text{av}}^{\text{p}}(\veR,\epsilon),
\end{align}
where we have defined
\begin{align}
    \left<\frac{1}{\tau(\ve{p}-\ve{q}_{\text{F}})}\right>_{q_{\text{F}}} = 2\pi n N_0 \left<|u(\ve{p}-\ve{q}_{\text{F}})|^2 \right>_{q_{\text{F}}}.
\end{align}
Above, $n$ is the density of non-magnetic impurities.\\
2) When we combine one ordinary scattering and one spin-orbit scattering on the same non-magnetic impurity, the self-energy is
\begin{align}
    \check{\sigma}^{uu_{\text{so}}}(\veR,\ve{p},\epsilon) =&-\frac{i\alpha q_{\text{F}}}{12}\left<\frac{1}{\tau(\vep-\veq_{\text{F}})}\right>_{q_{\text{F}}}\nabla_{\veR}\cdot[\hrhot\vehsigma\comtimes\check{\ve{g}}_{\text{av}}^{\text{p}}(\veR,\epsilon)]\notag\\
    &+\frac{i\alpha}{4}\left<\frac{1}{\tau(\vep-\veq_{\text{F}})}\right>_{q_{\text{F}}}\{\hrhot\vehsigma\times\ve{p}\comdot\nabla_{\veR}\check{g}^{\text{s}}_{\text{av}}(\veR,\epsilon)\}\notag\\
    &-\frac{\alpha q_{\text{F}}}{6}\left<\frac{1}{\tau(\vep-\veq_{\text{F}})}\right>_{q_{\text{F}}}\{\hrhot\vehsigma\times\ve{p}\comdot\check{\ve{g}}^{\text{p}}_{\text{av}}(\veR,\epsilon)\}.
\end{align}
3) For two spin-orbit scatterings to the second order in the impurity potential, we get
\begin{align}
    \check{\sigma}^{u_{\text{so}}^2}(\veR,\vep,\epsilon) = -\frac{i\alpha^2 q_{\text{F}}^2}{6}\left<\frac{1}{\tau(\vep-\veq_{\text{F}})}\right>_{q_{\text{F}}}[p^2\hrhot\vehsigma\cdot\check{g}^{\text{s}}_{\text{av}}(\veR,\epsilon)\hrhot\vehsigma
    -(\vep\cdot\hrhot\vehsigma)\check{g}^{\text{s}}_{\text{av}}(\veR,\epsilon)(\vep\cdot\hrhot\vehsigma)].
\end{align}
4) For ordinary scattering to the third order in the scattering potential, we get
\begin{align}
    \check{\sigma}^{u^3}(\veR,\vep,\epsilon) = \frac{1}{2\tau_{\text{sk}}(\vep)}\check{\rho}_0 ,
\end{align}
where we have defined
\begin{align}
    \frac{1}{\tau_{\text{sk}}(\vep)} = 2n(\pi N_0 )^2 \left<u(\vep-\veq_{\text{F}})u(\veq_{\text{F}}-\veq'_{\text{F}})u(\veq'_{\text{F}}-\vep)\right>_{q_{\text{F}},q'_{\text{F}}}.
\end{align}
5) For two ordinary impurity scatterings and one spin-orbit impurity scattering on the same non-magnetic impurity, we get
\begin{align}
    \check{\sigma}^{u^2 u_{\text{so}}}(\veR,\vep,\epsilon) = & 
    -\frac{\alpha}{4\tau_{\text{sk}}(\vep)}\vep\cdot\{\hrhot\vehsigma\comtimes\check{g}_{\text{av}}^{\text{s}}(\veR,\epsilon)\nabla_{\veR}\check{g}_{\text{av}}^{\text{s}}(\veR,\epsilon)\}\notag\\
    &-\frac{i\alpha p_{\text{F}}}{6\tau_{\text{sk}}(\vep)}\vep\cdot[\hrhot\vehsigma\comtimes\check{g}_{\text{av}}^{\text{s}}(\veR,\epsilon)\check{\ve{g}}_{\text{av}}^{\text{p}}(\veR,\epsilon)]\notag\\
    &-\frac{\alpha p_{\text{F}}}{12\tau_{\text{sk}}(\vep)}\Big\{\hrhot\vehsigma\cdot[\nabla_{\veR}\times\check{\ve{g}}_{\text{av}}^{\text{p}}(\veR,\epsilon)]\check{g}_{\text{av}}^{\text{s}}(\veR,\epsilon)
    +\frac{1}{2}\check{g}_{\text{av}}^{\text{s}}(\veR,\epsilon)\hrhot\vehsigma\cdot[\nabla_{\veR}\times\check{\ve{g}}_{\text{av}}^{\text{p}}(\veR,\epsilon)]\notag\\
    &\phantom{-\frac{\alpha p_{\text{F}}}{12\tau_{\text{sk}}(\vep)}}+\frac{1}{2}[\nabla_{\veR}\times\check{\ve{g}}_{\text{av}}^{\text{p}}(\veR,\epsilon)]\cdot\hrhot\vehsigma\check{g}_{\text{av}}^{\text{s}}(\veR,\epsilon)
    +\check{g}_{\text{av}}^{\text{s}}(\veR,\epsilon)[\nabla_{\veR}\times\check{\ve{g}}_{\text{av}}^{\text{p}}(\veR,\epsilon)]\cdot\hrhot\vehsigma\Big\}\notag\\
    &+\frac{\alpha p_{\text{F}}}{24\tau_{\text{sk}}(\vep)}\{\check{\ve{g}}_{\text{av}}^{\text{p}}(\veR,\epsilon)\cdot[\hrhot\vehsigma\times\nabla_{\veR}\check{g}_{\text{av}}^{\text{s}}(\veR,\epsilon)]-[\nabla_{\veR}\check{g}_{\text{av}}^{\text{s}}(\veR,\epsilon)\times\hrhot\vehsigma]\cdot\check{\ve{g}}_{\text{av}}^{\text{p}}(\veR,\epsilon)\}\notag\\
    &+\frac{i\alpha p_{\text{F}}^2}{18\tau_{\text{sk}}(\vep)}\check{\ve{g}}_{\text{av}}^{\text{p}}(\veR,\epsilon)\cdot[\hrhot\vehsigma\times\check{\ve{g}}_{\text{av}}^{\text{p}}(\veR,\epsilon)]\notag\\
    &-\frac{i\alpha}{8\tau_{\text{sk}}(\vep)}\Big([\nabla_{\veR}\check{g}_{\text{av}}^{\text{s}}(\veR,\epsilon)]\cdot\nabla_{\vep}\{\vep\cdot[\nabla_{\veR}\check{g}_{\text{av}}^{\text{s}}(\veR,\epsilon)\times\hrhot\vehsigma]\}\notag\\
    &\phantom{\frac{i\alpha}{8\tau_{\text{sk}}(\vep)}\Big(}+\nabla_{\vep}\{\vep\cdot[\hrhot\vehsigma\times\nabla_{\veR}\check{g}_{\text{av}}^{\text{s}}(\veR,\epsilon)]\}\cdot[\nabla_{\veR}\check{g}_{\text{av}}^{\text{s}}(\veR,\epsilon)]\Big)\notag\\
    &-\frac{\alpha p_{\text{F}}}{12\tau_{\text{sk}}(\vep)}\Big([\nabla_{\veR}\check{g}_{\text{av}}^{\text{s}}(\veR,\epsilon)]\cdot\nabla_{\vep}\{\vep\cdot[\check{\ve{g}}_{\text{av}}^{\text{p}}(\veR,\epsilon)\times\hrhot\vehsigma]\}\notag\\
    &\phantom{\frac{i\alpha}{8\tau_{\text{sk}}(\vep)}\Big(}-\nabla_{\vep}\{\vep\cdot[\hrhot\vehsigma\times\check{\ve{g}}_{\text{av}}^{\text{p}}(\veR,\epsilon)]\}\cdot[\nabla_{\veR}\check{g}_{\text{av}}^{\text{s}}(\veR,\epsilon)]\Big).
\end{align}
6) For two spin-flip scatterings on the same magnetic impurity, we get
\begin{align}
    \check{\sigma}^{u_{\text{sf}}^2}(\veR,\vep,\epsilon) = -\frac{iS(S+1)}{6}\left<\frac{1}{\tau_{\text{m}}(\vep-\veq_{\text{F}})}\right>_{q_{\text{F}}}\vehsigma\cdot\check{g}_{\text{av}}^{\text{s}}(\veR,\epsilon)\vehsigma,
\end{align}
where $S$ is the spin of the magnetic impurities, and
\begin{align}
    \left<\frac{1}{\tau_{\text{m}}(\vep-\veq_{\text{F}})}\right>_{q_{\text{F}}} = 2\pi n_{\text{m}} N_0 \left<|u_{\text{m}}(\ve{p}-\ve{q}_{\text{F}})|^2 \right>_{q_{\text{F}}}.
\end{align}
Above, $n_{\text{m}}$ is the density of magnetic impurities. To obtain the expression for the spin-flip scattering self-energy, we have averaged over all directions of the magnetic moments.

Now that we have found expressions for all of the self-energies, we can proceed to evaluate the odd and even equations. The odd equation gives rise to the expression for the current matrix of the system. A common assumption is that the scattering on non-magnetic impurities dominates over all other terms. Using the normalization condition
\begin{align}
    \check{g}_{\text{av}}(\veR,p_{\text{F}},\epsilon)\check{g}_{\text{av}}(\veR,p_{\text{F}},\epsilon)=\check{\rho}_0 
\end{align}
when evaluating the odd equation, we can express $\check{\ve{g}}_{\text{av}}^{\text{p}}(\veR,\epsilon)$ in terms of $\check{g}_{\text{av}}^{\text{s}}(\veR,\epsilon)$ as 
$\check{\ve{g}}_{\text{av}}^{\text{p}}(\veR,\epsilon)= -\tau v_{\text{F}} \check{g}_{\text{av}}^{\text{s}}(\veR,\epsilon)\nabla_{\veR}\check{g}_{\text{av}}^{\text{s}}(\veR,\epsilon).$
Within this approximation, this expression is proportional to the current matrix in the Usadel equation, which can be seen when inserting it into the even equation.
We want to include corrections to this result by including terms to the first order in the spin-orbit scattering strength. We assume that  
\begin{align}
    \check{\ve{g}}_{\text{av}}^{\text{p}}(\veR,\epsilon)= -\tau v_{\text{F}} \check{g}_{\text{av}}^{\text{s}}(\veR,\epsilon)\nabla_{\veR}\check{g}_{\text{av}}^{\text{s}}(\veR,\epsilon)+\delta\check{\ve{g}}^{\text{p}}_{\text{av}}(\veR,\epsilon),
    \label{eq:gp_expression_adjusted}
\end{align}
and insert this back into the odd equation. We assume that $|\check{\ve{g}}_{\text{av}}^{\text{p}}(\veR,\epsilon) |\ll \check{g}_{\text{av}}^{\text{s}}(\veR,\epsilon)$ and neglect terms of second order in $\check{\ve{g}}_{\text{av}}^{\text{p}}(\veR,\epsilon)$ as well as terms with one $\check{\ve{g}}_{\text{av}}^{\text{p}}(\veR,\epsilon)$ and one $\nabla_{\veR}\check{g}_{\text{av}}^{\text{s}}(\veR,\epsilon)$. We arrive at a correction
\begin{align}
    \delta\check{\ve{g}}^{\text{p}}_{\text{av}}(\veR,\epsilon) &= \left(\frac{\alpha p_{\text{F}}}{2}-\frac{\alpha v_{\text{F}}p_{\text{F}}^2 \tau^2}{3\tau_{\text{sk}}}\right)[\hrhot\vehsigma\comtimes\nabla_{\veR}\check{g}^{\text{s}}_{\text{av}}(\veR,\epsilon)]\notag\\
    &-\left(\frac{i\alpha v_{\text{F}}p_{\text{F}}^2 \tau}{3}+\frac{i\alpha p_{\text{F}}\tau}{2\tau_{\text{sk}}}\right)\{\hrhot\vehsigma\comtimes\check{g}^{\text{s}}_{\text{av}}(\veR,\epsilon)\nabla_{\veR}\check{g}^{\text{s}}_{\text{av}}(\veR,\epsilon)\}.
\end{align}
Above, we have introduced
\begin{align}
    1/\tau= & 2\pi n N_0 \big<|u(\ve{e}_{p_{\text{F}}}-\ve{e}_{q_{\text{F}}})|^2 \big>_{p_{\text{F}} ,q_{\text{F}}},\\
    1/\tau_{\text{sk}} =& 2\pi^2 n N_0^2 \big<u(\ve{e}_{p_{\text{F}}}-\ve{e}_{q_{\text{F}}})u(\ve{e}_{q_{\text{F}}}-\ve{e}_{q'_{\text{F}}})
    u(\ve{e}_{q'_{\text{F}}}-\ve{e}_{p_{\text{F}}}) \big>_{p_{\text{F}} ,q_{\text{F}},q'_{\text{F}}}.
\end{align}
Note that the requirement $\{\delta \check{g}_{\text{av}}^{\text{p}}(\veR,\epsilon),\check{g}^{\text{s}}_{\text{av}}(\veR,\epsilon)\}=0$ ensures that the Green's function follows the normalization condition.
The above expression for $\check{\ve{g}}_{\text{av}}^{\text{p}}(\veR,\epsilon)$ is inserted into the even equation to arrive at the Usadel equation.

The final step is to evaluate the even equation. The contribution from the odd equation enters the even equation through the term 
\begin{align}
    \frac{iv_{\text{F}}}{3}\nabla_{\veR}\cdot\check{\ve{g}}_{\text{av}}^{\text{p}}(\veR,\epsilon) = & -\frac{iv_{\text{F}}^2 \tau}{3}\nabla_{\veR}\cdot[\check{g}^{\text{s}}_{\text{av}}(\veR,\epsilon)\nabla_{\veR}\check{g}^{\text{s}}_{\text{av}}(\veR,\epsilon)]\notag\\
    &+\left(\frac{i\alpha v_{\text{F}}p_{\text{F}}}{6}-\frac{i\alpha v_{\text{F}}^2 p_{\text{F}}^2 \tau^2}{9\tau_{\text{sk}}}\right)\nabla_{\veR}\cdot[\hrhot\vehsigma\comtimes\nabla_{\veR}\check{g}_{\text{av}}^{\text{s}}(\veR,\epsilon)]\notag\\
    &+\left(\frac{\alpha v_{\text{F}}^2 p_{\text{F}}^2 \tau}{9}+\frac{\alpha v_{\text{F}}p_{\text{F}}\tau}{6\tau_{\text{sk}}}\right)\nabla_{\veR}\cdot\{\hrhot\vehsigma\comtimes\check{g}_{\text{av}}^{\text{s}}(\veR,\epsilon)\nabla_{\veR}\check{g}_{\text{av}}^{\text{s}}(\veR,\epsilon)\}.
    \label{eq:contribution_odd}
\end{align}
Because we have added extra $\mathcal{O}(\alpha^1 )$ terms containing a gradient of the Green's function in the odd equation, we need to include terms up to order $\mathcal{O}\{\nabla_{\veR}\check{g}_{\text{av}}^{\text{s}}(\veR,\epsilon)\times\nabla_{\veR}\check{g}_{\text{av}}^{\text{s}}(\veR,\epsilon) \}$ in the gradients for the $\mathcal{O}(\alpha^0 )$ and $\mathcal{O}(\alpha^1 )$ terms in the even equation. This is the reason why we need to consider first order terms in the gradient approximation in the even equation \cite{huang_prb_18}. Moreover, we assume that $l_{(\text{sk})} = \sqrt{D\tau_{(\text{sk})}}\gg p_{\text{F}}^{-1}$, which implies that $v_{\text{F}}p_{\text{F}}\gg \tau_{(\text{sk})}^{-1}$. This allows us to neglect terms that goes like $\alpha/\tau_{(\text{sk})}$. The contributions from the self-energy terms in the even equation in Eq.~\eqref{eq:even} are:
\newline
1) There is no contribution from $\check{\sigma}^{u^2}(\veR,\vep,\epsilon)$.
\newline
2) The contribution from $\check{\sigma}^{uu_{\text{so}}}(\veR,\vep,\epsilon)$ is 
\begin{align}
    &\frac{i\alpha v_{\text{F}}p_{\text{F}}}{6}\nabla_{\veR}\cdot[\hrhot\vehsigma\comtimes\nabla_{\veR}\check{g}_{\text{av}}^{\text{s}}(\veR,\epsilon)]
    +\frac{\alpha v_{\text{F}}^2 p_{\text{F}}^2 \tau}{18}[\hrhot\vehsigma\comdot[\nabla_{\veR}\check{g}_{\text{av}}^{\text{s}}(\veR,\epsilon)]\times[\nabla_{\veR}\check{g}_{\text{av}}^{\text{s}}(\veR,\epsilon)]]\notag\\
    &+\frac{i\alpha v_{\text{F}} p_{\text{F}} \tau}{6}[\hrhot\vehsigma\comdot\check{g}_{\text{av}}^{\text{s}}(\veR,\epsilon)[\nabla_{\veR}\check{g}_{\text{av}}^{\text{s}}(\veR,\epsilon)]\times[\nabla_{\veR}\check{g}_{\text{av}}^{\text{s}}(\veR,\epsilon)]]
\end{align}
3) The contribution from $\check{\sigma}^{u_{\text{so}}^2}(\veR,\vep,\epsilon)$ is
\begin{align}
    \frac{i\alpha^2 p_{\text{F}}^4}{9\tau}[\hrhot\vehsigma\cdot\check{g}_{\text{av}}^{\text{s}}(\veR,\epsilon)\hrhot\vehsigma,\check{g}_{\text{av}}^{\text{s}}(\veR,\epsilon)].
\end{align}
4) There is no contribution from $\check{\sigma}^{u^3}(\veR,\vep,\epsilon)$.
\newline
5) The contribution from $\check{\sigma}^{u^2u_{\text{so}}}(\veR,\vep,\epsilon)$ is
\begin{align}
    &\frac{\alpha v_{\text{F}}p_{\text{F}}\tau}{12\tau_{\text{sk}}}\nabla_{\veR}\cdot\{\hrhot\vehsigma\comtimes\check{g}_{\text{av}}^{\text{s}}(\veR,\epsilon)\nabla_{\veR}\check{g}_{\text{av}}^{\text{s}}(\veR,\epsilon)\}
    +\frac{\alpha v_{\text{F}} p_{\text{F}} \tau}{8\tau_{\text{sk}}}[\hrhot\vehsigma\comdot[\nabla_{\veR}\check{g}_{\text{av}}^{\text{s}}(\veR,\epsilon)]\times[\nabla_{\veR}\check{g}_{\text{av}}^{\text{s}}(\veR,\epsilon)]]\notag\\
    &-\frac{i\alpha v_{\text{F}}^2 p_{\text{F}}^2 \tau^2}{18\tau_{\text{sk}}}[\hrhot\vehsigma\comdot\check{g}_{\text{av}}^{\text{s}}(\veR,\epsilon)[\nabla_{\veR}\check{g}_{\text{av}}^{\text{s}}(\veR,\epsilon)]\times[\nabla_{\veR}\check{g}_{\text{av}}^{\text{s}}(\veR,\epsilon)]].
\end{align}
6) The contribution from $\check{\sigma}^{u_{\text{sf}}^2}(\veR,\vep,\epsilon)$ is
\begin{align}
    \frac{iS(S+1)}{18\tau_{\text{m}}}[\vehsigma\cdot\check{g}_{\text{av}}^{\text{s}}(\veR,\epsilon)\vehsigma,\check{g}_{\text{av}}^{\text{s}}(\veR,\epsilon)],
\end{align}
where we defined
\begin{align}
    1/\tau_{\text{m}}= & 2\pi n_{\text{m}} N_0 \big<|u_{\text{m}}(\ve{e}_{p_{\text{F}}}-\ve{e}_{q_{\text{F}}})|^2 \big>_{p_{\text{F}} ,q_{\text{F}}}.
\end{align}
Adding up all terms, we arrive at the Usadel equation
\begin{align}
\nabla_{\veR}\cdot\check{\ve{\mathcal{I}}}(\veR,\epsilon) = i[\check{\sigma}(\veR,\epsilon),\check{g}_{\text{av}}^{\text{s}}(\veR,\epsilon)]+\check{\mathcal{T}}(\veR,\epsilon).\label{eq:usadel}
\end{align}
Above, $\check{\ve{\mathcal{I}}}(\veR,\epsilon)=\check{\ve{\mathcal{I}}}^{(0)}(\veR,\epsilon)+\check{\ve{\mathcal{I}}}^{(1)}(\veR,\epsilon)$ is the current operator with zeroth and first order contributions
\begin{align}
    \check{\ve{\mathcal{I}}}^{(0)}(\veR,\epsilon) &=-D\check{g}_{\text{av}}^{\text{s}}(\veR,\epsilon)\nabla_{\veR}\check{g}_{\text{av}}^{\text{s}}(\veR,\epsilon),\label{eq:I0}\\
    \check{\ve{\mathcal{I}}}^{(1)}(\veR,\epsilon) &=D\Big(\frac{i\kappa}{2}\left\{\hrhot\vehsigma\comtimes\check{g}_{\text{av}}^{\text{s}}(\veR,\epsilon)\nabla_{\veR}\check{g}_{\text{av}}^{\text{s}}(\veR,\epsilon)\right\} +\frac{\theta}{2}\left[\hrhot \vehsigma \comtimes\nabla_{\veR}\check{g}_{\text{av}}^{\text{s}}(\veR,\epsilon)\right]\Big),\label{eq:I1}
\end{align}
with respect to the spin-orbit parameter $\alpha$, and 
\begin{align}
    \check{\mathcal{T}}(\veR,\epsilon) &=- \frac{D\theta}{4}\left[\hrhot\vehsigma\comdot\check{g}_{\text{av}}^{\text{s}}(\veR,\epsilon)(\nabla_{\veR}\check{g}_{\text{av}}^{\text{s}}(\veR,\epsilon))\times(\nabla_{\veR}\check{g}_{\text{av}}^{\text{s}}(\veR,\epsilon))\right]
    -\frac{iD\kappa}{4}\left[\hrhot\vehsigma\comdot(\nabla_{\veR}\check{g}_{\text{av}}^{\text{s}}(\veR,\epsilon))\times(\nabla_{\veR}\check{g}_{\text{av}}^{\text{s}}(\veR,\epsilon))\right]
\end{align}
is the torque. By evaluating its Keldysh component, we see that the torque only contributes to transport when the retarded Green's function is spatially dependent. The spin-Hall and spin-swap angles in the normal-state are given by
\begin{align}
    \theta =& -\frac{2\alpha p_{\text{F}}^2 \tau}{3\tau_{\text{sk}}}+\frac{2\alpha p_{\text{F}}}{v_{\text{F}}\tau},
    \:\:\:\:\:
    \kappa = -\frac{2\alpha p_{\text{F}}^2}{3}-\frac{3\alpha p_{\text{F}}}{2v_{\text{F}}\tau_{\text{sk}}}.
\end{align} 
The self-energy matrix in the Usadel equation is given by
\begin{align}
    \check{\sigma}(\veR,\epsilon)&=\hat{\sigma}_{\text{ssSC}}(\epsilon)+\hat{\sigma}_{\text{so}}(\veR,\epsilon)+\hat{\sigma}_{\text{sf}}(\veR,\epsilon)+\check{\sigma}_{\text{isct}}(\epsilon)
\end{align}
Above, $\hat{\sigma}_{\text{ssSC}}(\epsilon)=\epsilon\hrhot + \hat{\Delta}-\hat{\ve{\sigma}}\cdot\ve{m}$ is the self-energy of a spin-split superconductor, $\hat{\sigma}_{\text{so}}(\veR,\epsilon)  = \frac{i}{8\tau_{\text{so}}}\hrhot\vehsigma\cdot\check{g}_{\text{av}}^{\text{s}}(\veR,\epsilon)\hrhot\vehsigma$ describes spin-orbit scattering, $\hat{\sigma}_{\text{sf}}(\veR,\epsilon) = \frac{i}{8\tau_{\text{sf}}}\vehsigma\cdot\check{g}_{\text{av}}^{\text{s}}(\veR,\epsilon)\vehsigma$ describes spin-flip scattering, and  $\check{\sigma}_{\text{isct}}(\epsilon)=i\delta\text{diag}(\hrhot,-\hrhot)+2i\delta\tanh\left(\frac{\epsilon}{2T}\right)\text{antidiag}(\hrhot,0)$ describes inelastic scattering.
The scattering times associated with spin-flip and spin-orbit scattering are given by 
\begin{align}
    1/\tau_{\text{sf}}&=8\pi n_{\text{m}} N_0 \big<|u_{\text{m}} (\ve{e}_{p_F}-\ve{e}_{q_F})|^2 \big>_{p_F ,q_F }S(S+1)/3,\\
    1/\tau_{\text{so}}&=8\alpha^2 p_F^4 /(9\tau).
\end{align} 
We have included inelastic scattering in the relaxation time approximation which can be thought of as a constant tunneling coupling to an infinite normal-metal reservoir. This is a simple way of modelling $e-e$ or $e-$phonon scattering which causes decay of energy modes in the material.

As an important test to our solution, we show that the Usadel equation can be written in a commutator form. We evaluate the divergence of the matrix current in Eqs.~\eqref{eq:I0} and~\eqref{eq:I1}, and find that
\begin{align}
    \nabla_{\veR}\cdot\check{\ve{\mathcal{I}}}(\veR,\epsilon) &= 
    \check{\mathcal{T}}(\veR,\epsilon)
    +\frac{1}{2}[D\nabla_{\veR}^2\check{g}_{\text{av}}^{\text{s}}(\veR,\epsilon)\notag\\
    &+\big(\nabla_{\veR}\cdot\check{\ve{\mathcal{I}}}^{(1)}(\veR,\epsilon)\big)\check{g}_{\text{av}}^{\text{s}}(\veR,\epsilon)
    +\frac{D\theta}{2}[\nabla_{\veR}\check{g}_{\text{av}}^{\text{s}}(\veR,\epsilon)]\cdot[\hrhot\vehsigma\times\nabla_{\veR}\check{g}_{\text{av}}^{\text{s}}(\veR,\epsilon)],\check{g}_{\text{av}}^{\text{s}}(\veR,\epsilon)].
\end{align}
By inserting the right hand side of this expression into the Usadel equation, we realize that that the torque terms cancel so that it ensures that the Usadel equation [Eq.~\eqref{eq:usadel}] can be written as a commutator with $\check{g}_{\text{av}}^{\text{s}}(\veR,\epsilon)$. 

\section{II.\:\:The kinetic equations}

\hypertarget{sec:kinetic}{In} order to calculate the currents and the non-equilibrium charge and spin accumulations, we must solve the kinetic equations. These are obtained by evaluating the Keldysh part of the Usadel equation derived in the previous section.
We first assume that the retarded Green's function is equal to its equilibrium value
\begin{align}
    \hat{g}_{\text{av}}^{\text{R}}(\epsilon) =
    \begin{pmatrix}
    g_{+}(\epsilon) & 0 & 0 & f_{\text{s}}(\epsilon)+f_{\text{t}}(\epsilon)\\
    0 & g_{-}(\epsilon) & -f_{\text{s}}(\epsilon)+f_{\text{t}}(\epsilon) & 0\\
    0 & f_{\text{s}}(\epsilon)-f_{\text{t}}(\epsilon) &  -g_{-}(\epsilon) & 0\\
    -f_{\text{s}}(\epsilon)-f_{\text{t}}(\epsilon) & 0 & 0 & -g_{+}(\epsilon)
    \end{pmatrix},
\end{align}
where
\begin{align}
    g_{\pm}(\epsilon) & = (\epsilon+i\delta\pm m )I^{\pm}(\epsilon),
    \:\:\:\:\:
    f_{\text{s}}(\epsilon)  = \frac{\Delta}{2}\left[I^{+}(\epsilon)+I^- (\epsilon)\right],
    \:\:\:\:\:
    f_{\text{t}}(\epsilon)  = \frac{\Delta}{2}\left[I^{+}(\epsilon)-I^- (\epsilon)\right]
\end{align}
are the spin-split ordinary retarded Green's functions, and the spin-singlet and spin-triplet parts of the anomalous retarded Green's function, respectively. 
Above, $m$ is the magnitude of the spin-splitting field $\ve{m}=m\ve{z}$, $\delta$ is small and related to the strength of the inelastic scattering, and
\begin{align}
    I^{\pm}(\epsilon)=\frac{\text{sgn}(\epsilon\pm m )}{\sqrt{(\epsilon+i\delta\pm m )^2 -|\Delta|^2}}\Theta((\epsilon\pm m )^2 -|\Delta|^2 )-\frac{i}{\sqrt{|\Delta|^2-(\epsilon+i\delta\pm m )^2 }}\Theta(|\Delta|^2-(\epsilon\pm m )^2  ).
\end{align}
Assuming an equilibrium retarded Green's function makes it possible to obtain analytic expressions. 
The retarded Green's function can vary in space if 1) quasi-particle currents are transformed into supercurrents, 2) the superconducting gap is suppressed close to the interface, and 3) the superconducting gap is suppressed by energy injection.
The first can be disregarded, because the injected spin and energy currents do not support such a conversion. The transversal charge currents can in principle be converted into supercurrents, which is why we assume the width of the spin-split superconductor to be much smaller than the length over which this conversion occurs.
The second stems from the proximity effect between the injector and the spin-split superconductor. In experiments, the proximity effect can be minimized by using a tunnel barrier rather than a metallic contact.
The third is a relevant issue given that we inject an energy current into the spin-split superconductor. The energy distribution has been shown to suppress superconductivity entirely at a critical applied spin-voltage \cite{Keizer_PRL_2006}. However, in this work inelastic scattering was not taken into account. By including inelastic scattering, superconductivity should survive up to higher spin-voltages due to the decay of the energy current. A suppression of the superconducting gap and anomalous Green's function close to the injector is likely,
but should not cause qualitative changes in the quasi-particle currents as a function of the spin-splitting field.

The aim of solving the kinetic equations is to determine the non-equilibrium properties of the Green's function. We relate the advanced and Keldysh Green's functions to the retarded one by
\begin{align}
    \hat{g}^{\text{A}}_{\text{av}}(\epsilon) & = -[\hrhot\hat{g}^{\text{R}}_{\text{av}}(\epsilon)\hrhot]^{\dagger},
    \:\:\:\:\:
    \hat{g}^{\text{K}}_{\text{av}}(\epsilon)  = \hat{g}^{\text{R}}_{\text{av}}(\epsilon)\hat{h}(\veR,\epsilon)-\hat{h}(\veR,\epsilon)\hat{g}^{\text{A}}_{\text{av}}(\epsilon),
\end{align}
where the distribution function matrix 
\begin{align}
    \hat{h}(\veR,\epsilon) = 
    \hrhoz h_{\text{L}}(\veR,\epsilon) +
    \hrhot h_{\text{T}}(\veR,\epsilon) +
    \sum_i \hat{\sigma}_i  h_{\text{LS}i} (\veR,\epsilon)+
    \sum_i \hrhot\hat{\sigma}_i  h_{\text{TS}i} (\veR,\epsilon)
\end{align}
describes the non-equilibrium energy distribution $h_{\text{L}}(\ve{R},\epsilon)$, charge distribution $h_{\text{T}}(\veR,\epsilon)$, spin-energy distribution $h_{\text{LS}i}(\veR,\epsilon)$, and spin distribution $h_{\text{TS}i}(\veR,\epsilon)$. These can be found by performing appropriate traces over the distribution function matrix $\hat{h}(\veR,\epsilon)$. Starting from the continuity equations for the different currents \cite{espedal_prb_17}, we can show that performing the corresponding traces on the Keldysh part of the matrix current $\hat{\ve{\mathcal{I}}}(\veR,\epsilon)$, we obtain the energy resolved energy current $\ve{j}_{\text{L}}(\veR,\epsilon)$, charge current $\ve{j}_{\text{T}}(\veR,\epsilon)$, spin-energy current $\ve{j}_{\text{LS}i}(\veR,\epsilon)$, and spin current $\ve{j}_{\text{TS}i}(\veR,\epsilon)$.
We separate these currents into a zeroth order and a first order contribution in the spin-orbit parameter $\alpha$.
The zeroth order currents are related to the distribution functions by
\begin{align}
    \ve{j}_{\text{L}}^{(0)}(\veR,\epsilon)& = -2\left[D_{\text{L}}(\epsilon)\nabla_{\veR}h_{\text{L}}(\veR,\epsilon)+D_{\text{TS}z} (\epsilon)\nabla_{\veR}h_{\text{TS}z} (\veR,\epsilon)\right],\\
    \ve{j}_{\text{TS}x}^{(0)}(\veR,\epsilon)& = -2\left[D_{\text{TS}x} (\epsilon)\nabla_{\veR}h_{\text{TS}x} (\veR,\epsilon)+D_{\text{TS}y} (\epsilon)\nabla_{\veR}h_{\text{TS}y} (\veR,\epsilon)\right],\\
    \ve{j}_{\text{TS}y}^{(0)}(\veR,\epsilon)& = -2\left[D_{\text{TS}x} (\epsilon)\nabla_{\veR}h_{\text{TS}y} (\veR,\epsilon)-D_{\text{TS}y} (\epsilon)\nabla_{\veR}h_{\text{TS}x} (\veR,\epsilon)\right],\\
    \ve{j}_{\text{TS}z}^{(0)}(\veR,\epsilon)& = -2\left[D_{\text{L}} (\epsilon)\nabla_{\veR}h_{\text{TS}z} (\veR,\epsilon)+D_{\text{TS}z} (\epsilon)\nabla_{\veR}h_{\text{L}} (\veR,\epsilon)\right].
\end{align}
The first order currents are transversal to the zeroth order currents. The relevant first order currents are expressed in terms of the zeroth order currents in the main text.
The coefficients are given by
\begin{align}
    D_{\text{L}}(\epsilon) &= 
    \frac{D}{2}\Big\{1+\frac{1}{2}\left[|g_+ (\epsilon)|^2 +|g_- (\epsilon)|^2 -2|f_{\text{s}}(\epsilon)|^2 -2|f_{\text{t}}(\epsilon)|^2 \right]\Big\}\\
    D_{\text{TS}x} (\epsilon) & =
    \frac{D}{2}\left\{1+\Real\{g_+ (\epsilon)[g_- (\epsilon)]^* \} -|f_{\text{s}}(\epsilon)|^2 +|f_{\text{t}}(\epsilon)|^2 \right\}\\
    D_{\text{TS}y} (\epsilon) & = 
    \frac{D}{2}\left(\Imag\{g_+ (\epsilon)[g_- (\epsilon)]^* \} +\Imag\{f_{\text{s}}(\epsilon) [f_{\text{t}}(\epsilon)]^* \} \right),\\
    D_{\text{TS}z} (\epsilon) & =
    \frac{D}{4}\left[|g_+ (\epsilon)|^2 -|g_- (\epsilon)|^2 -4\Real\{f_{\text{s}}(\epsilon) [f_{\text{t}}(\epsilon)]^* \} \right].
\end{align}
We also define
\begin{align}
    N_{\pm} (\epsilon) & = \{\Real[g_{+}(\epsilon)]\pm \Real[g_- (\epsilon)]\}/2,\\
    N_{-}^{\text{I}}(\epsilon) & = \{\Imag[g_{+}(\epsilon)]- \Imag[g_- (\epsilon)]\}/2
\end{align} 
in order to express the first order currents in the main text. $N_0N_+(\epsilon)$ is the density-of-states, where $N_0$ is the Fermi level density-of-states in the normal state.

We have now obtained analytic expressions for the first order currents in terms of the zeroth order ones. To evaluate the zeroth order currents, we must solve the kinetic equations for the injected energy and spin currents numerically. The kinetic equations are given by
\begin{align}
    \nabla_{\veR}\cdot \ve{j}_{\text{L}}^{(0)}(\veR,\epsilon)  = &
    -4\delta\left\{N_+ (\epsilon)\left[h_{\text{L}}(\veR,\epsilon)-\tanh\left(\frac{\epsilon}{2T}\right)\right]+N_- (\epsilon)h_{\text{TS}z} (\veR,\epsilon)\right\},\\
    \nabla_{\veR}\cdot \ve{j}_{\text{TS}z}^{(0)} (\veR,\epsilon)  = &-2\left[\frac{\alpha_{\text{so}}(\epsilon)}{\tau_{\text{so}}}+\frac{\alpha_{\text{sf}}(\epsilon)}{\tau_{\text{sf}}}\right]h_{\text{TS}z} (\veR,\epsilon)-4\delta\left\{N_+ (\epsilon)h_{\text{TS}z}(\veR,\epsilon)+N_- (\epsilon)\left[h_{\text{L}}(\veR,\epsilon)-\tanh\left(\frac{\epsilon}{2T}\right)\right]\right\},\\
    \nabla_{\veR}\cdot j_{\text{TS}x}^{(0)} (\veR,\epsilon)  = &
    4m [N_-^{\text{I}}(\epsilon)h_{\text{TS}x} (\veR,\epsilon)-N_+ (\epsilon)h_{\text{TS}y} (\veR,\epsilon)]\notag\\
    &-2\left[\frac{\alpha_{\text{so}}^x (\epsilon)}{\tau_{\text{so}}}+\frac{\alpha_{\text{sf}}^x (\epsilon)}{\tau_{\text{sf}}}\right]h_{\text{TS}x} (\veR,\epsilon)
    -2\left[\frac{\alpha_{\text{so}}^y (\epsilon)}{\tau_{\text{so}}}+\frac{\alpha_{\text{sf}}^y (\epsilon)}{\tau_{\text{sf}}}\right]h_{\text{TS}y} (\veR,\epsilon)\notag\\
    &-4\delta\left[N_+ (\epsilon)h_{\text{TS}x} (\veR,\epsilon) +N_-^{\text{I}}(\epsilon)h_{\text{TS}y} (\veR,\epsilon)\right],\\
    \nabla_{\veR}\cdot j_{\text{TS}y}^{(0)} (\veR,\epsilon)  = &
    4m [N_-^{\text{I}}(\epsilon)h_{\text{TS}y} (\veR,\epsilon)+N_+ (\epsilon)h_{\text{TS}x} (\veR,\epsilon)]\notag\\
    &-2\left[\frac{\alpha_{\text{so}}^x (\epsilon)}{\tau_{\text{so}}}+\frac{\alpha_{\text{sf}}^x (\epsilon)}{\tau_{\text{sf}}}\right]h_{\text{TS}y} (\veR,\epsilon)
    +2\left[\frac{\alpha_{\text{so}}^y (\epsilon)}{\tau_{\text{so}}}+\frac{\alpha_{\text{sf}}^y (\epsilon)}{\tau_{\text{sf}}}\right]h_{\text{TS}x} (\veR,\epsilon)\notag\\
    &-4\delta\left[N_+ (\epsilon)h_{\text{TS}y} (\veR,\epsilon) -N_-^{\text{I}}(\epsilon)h_{\text{TS}x} (\veR,\epsilon)\right], 
\end{align}
where
\begin{align}
    \alpha_{\text{so}}(\epsilon) &= 
    \Real[g_{\uparrow}(\epsilon)]\Real[g_{\downarrow}(\epsilon)]-\{\Real[f_{\text{s}}(\epsilon)]\}^2 + \{\Real[f_{\text{t}}(\epsilon)]\}^2 \\
    \alpha_{\text{sf}}(\epsilon) &= 
    \Real[g_{\uparrow}(\epsilon)]\Real[g_{\downarrow}(\epsilon)]+\{\Real[f_{\text{s}}(\epsilon)]\}^2 - \{\Real[f_{\text{t}}(\epsilon)]\}^2,\\
    \alpha_{\text{so}}^x (\epsilon) & = [N_+ (\epsilon)]^2 -\{\Real[f_{\text{s}}(\epsilon)]\}^2 ,\\
    \alpha_{\text{so}}^y (\epsilon) & = \left\{\Real[g_+ (\epsilon)]\Imag[g_+ (\epsilon)] - \Real[g_- (\epsilon)]\Imag[g_- (\epsilon)]- \Imag[g_+ (\epsilon) g_-^* (\epsilon)]-4\Real[f_{\text{s}}(\epsilon)]\Imag[f_{\text{t}}(\epsilon)]\right\}/4,\\
    \alpha_{\text{sf}}^x (\epsilon) & = \left(\{\Real[g_+ (\epsilon)] + \Real[g_- (\epsilon)]\}^2 + 4\{\Real[f_{\text{s}}(\epsilon)]\}^2 \right)/4,\\
    \alpha_{\text{sf}}^y (\epsilon) & = \left(\{\Imag[g_+ (\epsilon) + g_-^* (\epsilon)]\}^2 /4 + \Imag[f_{\text{s}}(\epsilon)f_{\text{t}} (\epsilon)]-\Imag[f_{\text{s}}(\epsilon)f_{\text{t}}^* (\epsilon)]\right)/2.
\end{align}
We want to consider the dependence on temperature and spin-splitting. To do so, we assume that the superconducting pairing is equal to its value in a uniform spin-split superconductor. The gap equation for such a bulk superconductor is given by
\begin{align}
    \Delta = \frac{N_0 U}{2}\int d\epsilon\:\Real[f_{\text{s}}(\epsilon)]\tanh\left(\frac{\epsilon}{2T}\right).
\end{align}
We solve this equation self-consistently with the Debye cutoff $\omega_{\text{D}}=\Delta_0 \cosh(1/N_0 U)$. In all figures, we use $\omega_{\text{D}}\approx74\Delta_0$.

To solve the kinetic equations, we must define boundary conditions.
We use the Kupriyanov-Lukichev boundary condition 
\begin{align}
    \ve{n}_{1\to2}\cdot 2L_j \zeta_j [\check{g}_{\text{av}}^{\text{s}}(\veR,\epsilon)]_j \nabla_{\veR}[\check{g}_{\text{av}}^{\text{s}}(\veR,\epsilon)]_j = [[\check{g}_{\text{av}}^{\text{s}}(\veR,\epsilon)]_1 ,[\check{g}_{\text{av}}^{\text{s}}(\veR,\epsilon)]_2],
\end{align}
where $\ve{n}_{1\to2}$ is the normal unit vector from material 1 to material 2, $L_{j}$ is the length of material $j\in\{1,2\}$ in the direction of the interface normal, and $\zeta_j = R_{\text{B}}/R_j$ is the ratio between the barrier resistance and the resistance of material $j$. In Fig.~3 in the main text, we have used $\zeta_{\text{SC}}=4$.
First consider a normal-metal/spin-split superconductor (NM/ssSC) interface at $x_i = 0$ where a $z$ polarized spin-voltage $V_{\uparrow}=-V_{\downarrow}=V/2$ is applied to the NM. This corresponds to the boundary conditions
\begin{align}
    j_{\text{L}}^{i(0)}(0^+ ,\epsilon) & = -\frac{D}{L_{\text{SC}}\zeta_{\text{SC}}}\{N_+ (\epsilon)[h_{\text{L}}(0^{+} ,\epsilon)-h_{\text{L}}(0^{-} ,\epsilon)] + N_- (\epsilon) [h_{\text{TS}z} (0^{+} ,\epsilon)-h_{\text{TS}z} (0^{-} ,\epsilon)]\},\\
    j_{\text{TS}z}^{i(0)}(0^{+} ,\epsilon) & = -\frac{D}{L_{\text{SC}}\zeta_{\text{SC}}}\{N_+ (\epsilon) [h_{\text{TS}z}(0^+ ,\epsilon)-h_{\text{TS}z}(0^- ,\epsilon)] + N_- (\epsilon) [h_{\text{L}} (0^+,\epsilon)-h_{\text{L}} (0^- ,\epsilon)]\},
\end{align}
with
\begin{align}
    h_{\text{L}}(0^- ,\epsilon) & = \frac{1}{2}\left[\tanh\left(\frac{\epsilon+eV/2}{2T}\right)+\tanh\left(\frac{\epsilon-eV/2}{2T}\right)\right],\label{eq:hL}\\
    h_{\text{TS}z}(0^- ,\epsilon) & = \frac{1}{2}\left[\tanh\left(\frac{\epsilon+eV/2}{2T}\right)-\tanh\left(\frac{\epsilon-eV/2}{2T}\right)\right]\label{eq:hTSz}.
\end{align}
If we instead consider an $x$ polarized spin-voltage, the boundary condition at $x_i=0$ is
\begin{align}
    j_{\text{L}}^{i(0)}(0^+ ,\epsilon) & = -\frac{D}{L_{\text{SC}}\zeta_{\text{SC}}}\{N_+ (\epsilon)[h_{\text{L}}(0^{+} ,\epsilon)-h_{\text{L}}(0^{-} ,\epsilon)] + N_- (\epsilon) h_{\text{TS}z} (0^{+} ,\epsilon)\},\\
    j_{\text{TS}x}^{i(0)}(0^+ ,\epsilon) & = -\frac{D}{L_{\text{SC}}\zeta_{\text{SC}}}\{N_+ (\epsilon)[h_{\text{TS}x}(0^+ ,\epsilon)-h_{\text{TS}x}(0^- ,\epsilon)] + N_-^{\text{I}} (\epsilon) h_{\text{TS}y} (0^+ ,\epsilon)\},\\
    j_{\text{TS}y}^{i(0)}(0^+ ,\epsilon) & = -\frac{D}{L_{\text{SC}}\zeta_{\text{SC}}}\{N_+ (\epsilon) h_{\text{TS}y}(0^+  ,\epsilon) - N_-^{\text{I}} (\epsilon) [h_{\text{TS}x} (0^+ ,\epsilon)-h_{\text{TS}x} (0^- ,\epsilon)]\},\\
    j_{\text{TS}z}^{i(0)}(0^{+} ,\epsilon) & = -\frac{D}{L_{\text{SC}}\zeta_{\text{SC}}}\{N_+ (\epsilon) h_{\text{TS}z}(0^+ ,\epsilon) + N_- (\epsilon) [h_{\text{L}} (0^+,\epsilon)-h_{\text{L}} (0^- ,\epsilon)]\},
\end{align}
with $h_{\text{L}}(0^-,\epsilon)$ following Eq.~\eqref{eq:hL}, and $h_{\text{TS}x}(0^-,\epsilon)$ being equal to $h_{\text{TS}z}(0^-,\epsilon)$ for the $z$ polarized case (Eq.~\eqref{eq:hTSz}).
For the right interface at $x_i = L_{\text{SC}}$, we consider a ssSC/vacuum interface where all currents are zero.

The relevant length scales should obey $l\ll l_{\text{so}},l_{\text{sf}}\ll l_{\text{isc}} < L_{\text{SC}}$. We have defined the scattering length for scattering on non-magnetic impurities as $l=\sqrt{D\tau}$ and the normal-state scattering length for spin-orbit (spin-flip) scattering as $l_{\text{so(sf)}}=\sqrt{D\tau_{\text{so(sf)}}}$. The normal-state inelastic scattering length is given by $l_{\text{isc}}=\sqrt{D/2\delta }$. Additionally, a precession length can be defined as $l_{\text{prec}}=\sqrt{D/2m}$ which is determined by the strength of the spin-splitting field.
In Fig.~3 in the main text, we use
$l_{\text{so}}=l_{\text{sf}}=20l$, $l_{\text{isc}}=250l$, and $L_{\text{SC}}=2l_{\text{isc}}$. We use that $l_{\text{prec}}=\sqrt{\delta/m}\:l_{\text{isc}}$.

\begin{figure}[t!]
    \centering
    \includegraphics[width=0.8\textwidth]{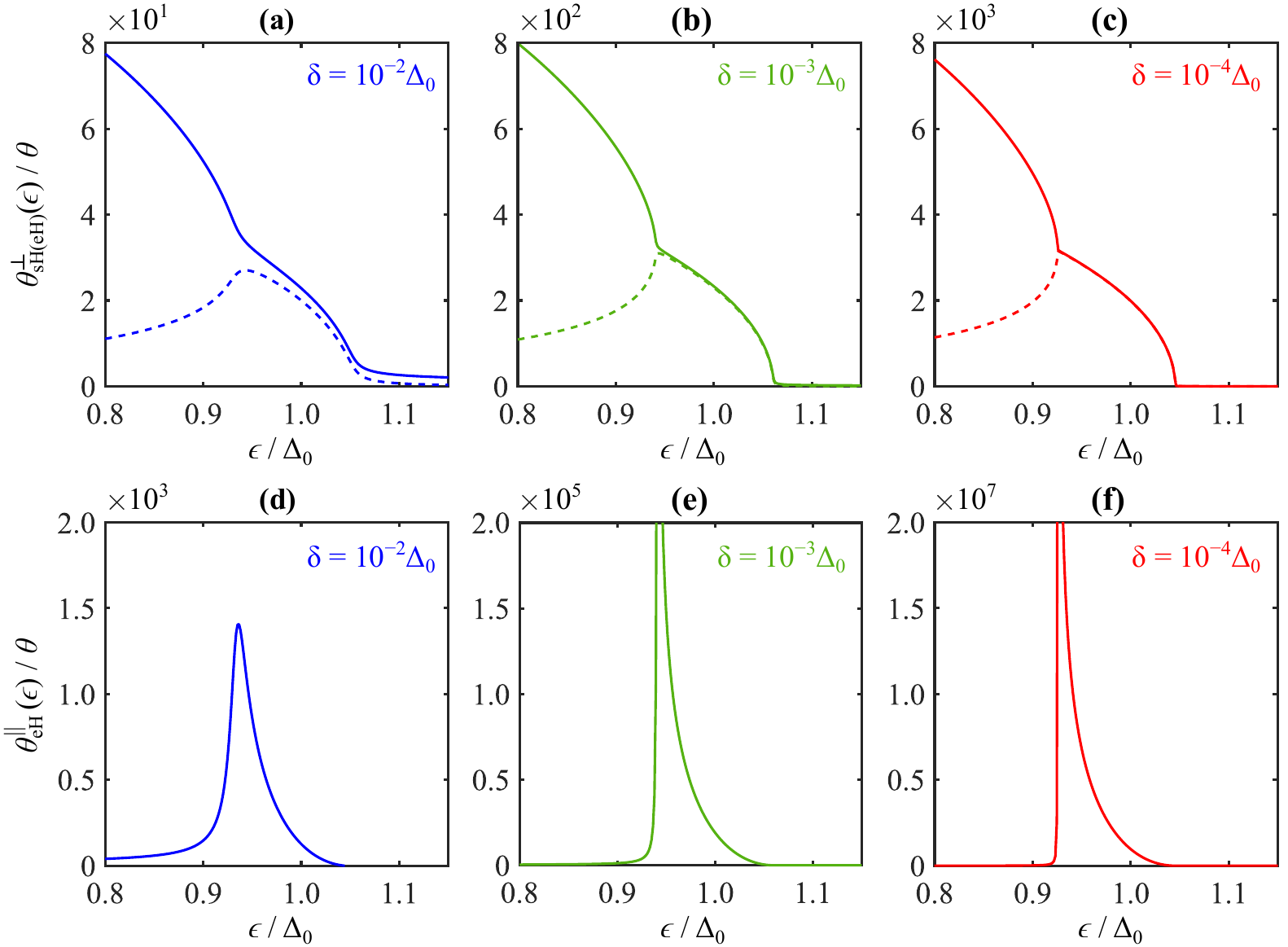}
    \caption{(a)-(c) The spin-Hall angle $\theta_{\text{sH}}^{\perp}$ (solid lines) and the energy-Hall angle $\theta_{\text{eH}}^{\perp}$ (dashed lines), and 
    (d)-(e) the energy-Hall angle $\theta_{\text{eH}}^{\parallel}$  for inelastic scattering parameter $\delta=10^{-2}\Delta_0$, $10^{-3}\Delta_0$ and $10^{-4}\Delta_0$ from left to right at spin-splitting field $m=0.06\Delta_0$.  
    We consider zero temperature, so that $\Delta=\Delta_0$ at $m=0$.}
    \label{fig:SI01}
\end{figure}

Experimentally, it is known that the range of $\delta/\Delta_0$ in the low-temperature regime $T\ll T_c$ of Al-superconductors can be of order $10^{-4}$ to $10^{-5}$ \cite{Feshchenko_PhysRevAppl_2015}. There are also experiments, where the data was fitted using a theoretical model with $\delta/\Delta_0=10^{-2}$ \cite{Strambini_NatCommun_2022}.
In the main text, we consider an inelastic scattering parameter $\delta = 10^{-3}\Delta_0$ that falls in the middle of these values and should be an experimentally realistic choice.
In Fig.~\ref{fig:SI01}, we furthermore provide results for the spin-Hall and energy-Hall angles $\theta_{\text{sH}}^{\perp}$, $\theta_{\text{eH}}^{\perp}$, and $\theta_{\text{eH}}^{\parallel}$ for additional values of $\delta/\Delta_0$ at a fixed magnetization.
The renormalization of the spin-Hall angle $\theta_{\text{sH}}^{\perp}$ and the energy-Hall angle $\theta_{\text{eH}}^{\perp}$ approximately depend on the inelastic scattering parameter through $(\delta/\Delta_0)^{-1}$, meaning that if we increase (decrease) $\delta/\Delta_0$ by one order of magnitude, $\theta_{\text{sH}}^{\perp}$ and $\theta_{\text{eH}}^{\perp}$ decrease (increase) by one order of magnitude.
The renormalization of the energy-Hall angle $\theta_{\text{eH}}^{\parallel}$ approximately depends on the inelastic scattering parameter through $(\delta/\Delta_0)^{-2}$, meaning that if we increase (decrease) $\delta/\Delta_0$ by one order of magnitude, $\theta_{\text{eH}}^{\parallel}$ decreases (increases) by two orders of magnitude. 
The latter also holds for the spin-swap angle $\kappa_{\text{es}}$.
Thus, an even larger renormalization of the spin-Hall, energy-Hall and spin-swap angles than what is presented in Fig.~2 in the main text can be achieved by choosing materials where $\delta/\Delta_0$ is small.

In the present work, we consider inelastic scattering lengths shorter than the length of the superconductor ($l_{\text{isc}}<L_{\text{SC}}$) so that all currents have decayed to zero before the superconductors right interface.
If $l_{\text{isc}}>L_{\text{SC}}$, back-flow currents would ensure the validity of the boundary condition of zero current through the interface towards vacuum. In this case, the decay length of the signal would be restricted by $L_{\text{SC}}$ in addition to $l_{\text{isc}}$.
If the right end of the superconductor is connected to an unbiased normal-metal instead of vacuum, back-flow currents are reduced since the energy current can penetrate this region. Thus, the inverse spin-Hall and spin-swapping signals can be measured closer to the right interface.

\section{III.\:\:The non-equilibrium charge and spin accumulations}

\hypertarget{sec:non-equilibrium_accumulations}{After} solving the kinetic equations with a self-consistently determined gap $\Delta$ with the Kupriyanov-Lukichev boundary conditions described in the previous section, we can calculate the non-equilibrium charge and spin accumulations. In a spin-split superconductor, the charge accumulation $\mu(x_j)$, and the $i$ polarized spin accumulations $\mu^{\text{s}}_i(x_j)$ are given by
\begin{align}
    \frac{\mu(x_j )}{e} &=-\frac{1}{2}\int d\epsilon\:\{N_{+}(\epsilon)h_{\text{T}}(x_j ,\epsilon)+N_{-}(\epsilon)h_{\text{LS}z} (x_j ,\epsilon)\},\\
    \frac{\mu_x^{\text{s}}(x_j )}{1/2} &=-\frac{1}{2}\int d\epsilon\:\{N_{+}(\epsilon)h_{\text{TS}x} (x_j ,\epsilon)+N^{\text{I}}_{-}(\epsilon)h_{\text{TS}y} (x_j ,\epsilon)\},\\
    \frac{\mu_y^{\text{s}}(x_j )}{1/2} &=-\frac{1}{2}\int d\epsilon\:\{N_{+}(\epsilon)h_{\text{TS}y} (x_j ,\epsilon)-N^{\text{I}}_{-}(\epsilon)h_{\text{TS}x} (x_j ,\epsilon)\},\\
    \frac{\mu_z^{\text{s}}(x_j )}{1/2} &=-\frac{1}{2}\int d\epsilon\:\left\{N_{+}(\epsilon)h_{\text{TS}z} (x_j ,\epsilon)+N_{-}(\epsilon)\left[h_{\text{L}} (x_j ,\epsilon)-\tanh\left(\frac{\epsilon}{2T}\right)\right]\right\},
\end{align}
for $i=x,y,z$.
We want to find the resulting charge and spin accumulation across the superconductor due to the transversal currents. If we measure the voltage or spin-voltage, the current going through the measuring circuit is negligible. We therefore require that there is no net transversal current at the edges of the sample. A current is induced between the edges of the sample to compensate for the first order transversal current. Our boundary condition is a complete cancellation of these currents at the edges.
We assume that the width $W$ of the superconductor is short enough that there is no conversion between quasi-particle and supercurrent, and that there is no significant scattering apart from the ordinary scattering on non-magnetic impurities.
We also assume that the coupling between the $x$ and $y$ polarized spin currents is weak so that the precession around the spin-splitting field is slow compared to the time it takes for the spin current to cross the width of the sample.
The relative charge accumulation $\Delta\mu(x_i) = \mu(x_i,W/2)-\mu(x_i,-W/2)$ across the spin-split superconductor is given by
\begin{align}
    \Delta\mu(x_i) =-\frac{W}{2}\int d\epsilon\:\bigg\{&\frac{N_{+}(\epsilon)D_{\text{T}}(\epsilon)-N_{-}(\epsilon)D_{\text{LS}z} (\epsilon)}{[D_{\text{T}}(\epsilon)]^2 -[D_{\text{LS}z}(\epsilon)]^2}j_{\text{T}}^{j(1)}(x_i ,\epsilon)
    -\frac{N_{+}(\epsilon)D_{\text{LS}z}(\epsilon)-N_{-}(\epsilon)D_{\text{T}}(\epsilon)}{[D_{\text{T}}(\epsilon)]^2 -[D_{\text{LS}z}(\epsilon)]^2}j_{\text{LS}z}^{j(1)}(x_i ,\epsilon)\bigg\},
\end{align}
where $x_i$ is the distance from the normal-metal contact, and
\begin{align}
    D_{\text{T}}(\epsilon) & = \frac{D}{2}\left\{1+\frac{1}{2}\left[|g_+ (\epsilon)|^2 +|g_- (\epsilon)|^2 +2|f_{\text{s}}(\epsilon)|^2 +2|f_{\text{t}}(\epsilon)|^2 \right]\right\},\\
    D_{\text{LS}z} (\epsilon) & =\frac{D}{4}\left[|g_+ (\epsilon)|^2 -|g_- (\epsilon)|^2 +4\Real\{f_{\text{s}}(\epsilon) [f_{\text{t}}(\epsilon)]^* \} \right].
\end{align}
The relative non-equilibrium spin accumulations $\Delta\mu^j_{\text{S}}(x_i) = \mu^j_{\text{S}}(x_i,W/2)-\mu^j_{\text{S}}(x_i,-W/2)$ across the spin-split superconductor are given by
\begin{align}
    \Delta\mu_{\text{S}}^x (x_i) =-\frac{W}{2}\int d\epsilon\:\bigg\{&\frac{N_{+}(\epsilon)D_{\text{TS}x} (\epsilon)+N_{-}^{\text{I}}(\epsilon)D_{\text{TS}y} (\epsilon)}{[D_{\text{TS}x} (\epsilon)]^2 +[D_{\text{TS}y}(\epsilon)]^2}j_{\text{TS}x}^{j(1)}(x_i ,\epsilon)\notag\\
    -&\frac{N_{+}(\epsilon)D_{\text{TS}y}(\epsilon)-N_{-}^{\text{I}}(\epsilon)D_{\text{TS}x} (\epsilon)}{[D_{\text{TS}x} (\epsilon)]^2 +[D_{\text{TS}y}(\epsilon)]^2}j_{\text{TS}y}^{j(1)}(x_i ,\epsilon)\bigg\},\\
    \Delta\mu_{\text{S}}^y(x_i) =-\frac{W}{2}\int d\epsilon\:\bigg\{&\frac{N_{+}(\epsilon)D_{\text{TS}x} (\epsilon)+N_{-}^{\text{I}}(\epsilon)D_{\text{TS}y} (\epsilon)}{[D_{\text{TS}x} (\epsilon)]^2 +[D_{\text{TS}y}(\epsilon)]^2}j_{\text{TS}y}^{j(1)}(x_i ,\epsilon)\notag\\
    +&\frac{N_{+}(\epsilon)D_{\text{TS}y}(\epsilon)-N_{-}^{\text{I}}(\epsilon)D_{\text{TS}x} (\epsilon)}{[D_{\text{TS}x} (\epsilon)]^2 +[D_{\text{TS}y}(\epsilon)]^2}j_{\text{TS}x}^{j(1)}(x_i ,\epsilon)\bigg\},\\
    \Delta\mu_{\text{S}}^z(x_i) =-\frac{W}{2}\int d\epsilon\:\bigg\{&\frac{N_{+}(\epsilon)D_{\text{L}}(\epsilon)-N_{-}(\epsilon)D_{\text{TS}z} (\epsilon)}{[D_{\text{L}}(\epsilon)]^2 -[D_{\text{TS}z}(\epsilon)]^2}j_{\text{L}}^{j(1)}(x_i ,\epsilon)
    -\frac{N_{+}(\epsilon)D_{\text{TS}z}(\epsilon)-N_{-}(\epsilon)D_{\text{L}}(\epsilon)}{[D_{\text{L}}(\epsilon)]^2 -[D_{\text{TS}z}(\epsilon)]^2}j_{\text{TS}z}^{j(1)}(x_i ,\epsilon)\bigg\}.
\end{align}
We have absorbed a normalization factor $e$ into the relative charge accumulation, and a factor $1/2$ into the relative spin accumulations for simplicity of notation.

\section{IV.\:\:Inverse spin-Hall signal for different applied spin-voltages}

\begin{figure}[h]
    \centering
    \includegraphics[width=0.5\textwidth]{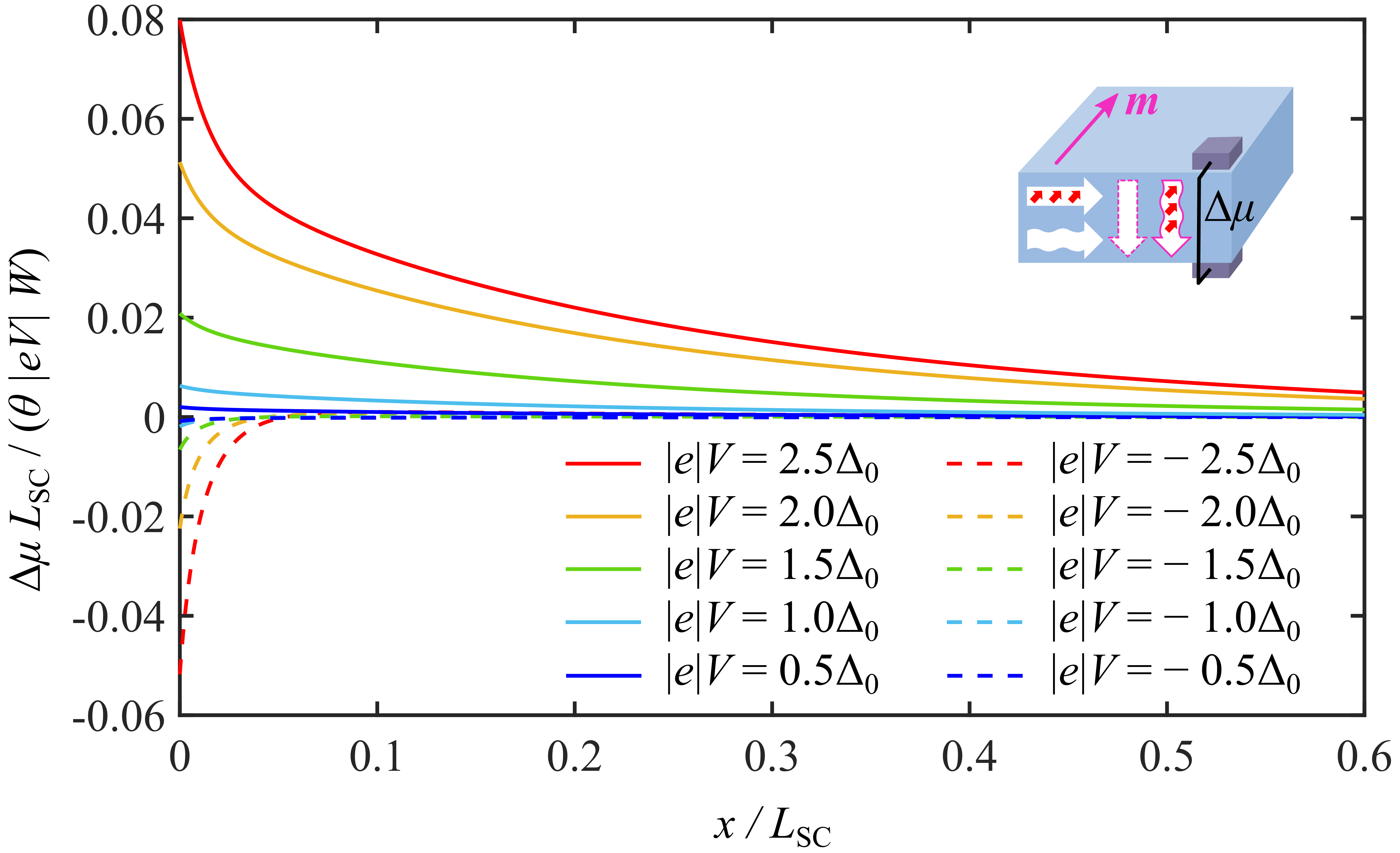}
    \caption{The out-of-plane charge accumulation in the spin-split superconductor ($\boldsymbol{m}\parallel$ spin) in Fig.~3(a) in the main text for different spin-voltages. The charge accumulation is normalized by $\theta|eV|W/L_{\text{SC}}$, where $\theta$ is the normal-state spin-Hall angle, $e$ is the electron charge, $V =(V_{\uparrow}-V_{\downarrow})$ is the spin-voltage in the injector, $W$ is the distance between the detectors, and $L_{\text{SC}}$ is the length of the ssSC. We consider $T = T_c/4$, and a spin-splitting field of magnitude $m = 0.1\Delta_0$, where $\Delta_0$ is the zero temperature gap at $m = 0$.}
    \label{fig:voltages}
\end{figure}

\hypertarget{sec:voltages}{As} mentioned previously, the energy distribution has been shown to suppress superconductivity entirely at a critical applied spin-voltage \cite{Keizer_PRL_2006}. We assume the superconducting gap to be constant in space, thus disregarding this suppression. However, by including inelastic scattering, superconductivity should survive up to higher spin-voltages due to the decay of the energy current away from the interface. In Fig.~\ref{fig:voltages}, we show the out-of-plane charge accumulation, arising due to the inverse spin Hall effect, as a function of the distance from the interface for a spin-split superconductor with the spin-splitting field parallel to the injected spin at different spin-voltages. We have used the same parameters as in Fig.~3(a) in the main text. We find a significant charge accumulation compared to the spin-voltage, also for lower positive spin-voltages ($V>0$). For negative spin-voltages ($V<0$), the inverse spin-Hall signal from the energy injection and spin injection act destructively, and the contribution from the energy current is small.

\twocolumngrid



%

\end{document}